\documentclass[11pt,onecolumn]{article}
\advance\textheight by \topskip
\usepackage{fixltx2e,fix-cm}
\usepackage{amssymb}
\usepackage{amsmath}
\usepackage{graphicx}
\usepackage{subfigure}
\usepackage{makeidx}
\usepackage{multicol}
\usepackage{needspace}
\usepackage{lipsum}
\usepackage{url}
\usepackage{xparse}% http://ctan.org/pkg/xparse
\NewDocumentCommand{\rot}{O{45} O{1em} m}{\makebox[#2][l]{\rotatebox{#1}{#3}}}%
\usepackage{booktabs}

\makeindex

\graphicspath{ {figures/} }

\newcommand{\qc}{\c{c}}

\begin{document}

\title{Extreme-Scale De Novo Genome Assembly  \footnote{To appear as a chapter in \emph{Exascale Scientific Applications: Programming Approaches for Scalability, Performance, and Portability, Straatsma, Antypas, Williams (editors), CRC Press, 2017}}} 

\author{\vspace{0.2cm} Evangelos Georganas$^1$, Steven Hofmeyr$^2$, Rob Egan$^3$, Ayd\i n Bulu\qc$^2$, \\
Leonid Oliker$^2$, Daniel Rokhsar$^3$, Katherine Yelick$^2$ \\
\newline \\
{\em  $^1$Intel Corporation, Santa Clara, USA} \\
{\em $^2$Joint Genome Institute, Lawrence Berkeley National Laboratory, Berkeley, USA}\\
{\em $^3$Computational Research Division, Lawrence Berkeley National Laboratory, Berkeley, USA}\\
}
\date{}

%\frontmatter

\maketitle%This is a placeholder titlepage, it will not be final.

%\include{frontmatter/dedication}
%\cleardoublepage
%\setcounter{page}{7} %previous pages reserved for frontmatter to be added later
%\tableofcontents
%\listoffigures
%\listoftables
%\include{frontmatter/foreword}
%\include{frontmatter/preface}
%\include{frontmatter/contributor}

%\mainmatter

%\part{This is a Part}
De novo whole genome assembly reconstructs genomic sequence from short, overlapping, and potentially erroneous DNA segments and is one of the most important computations in modern genomics. This work presents HipMER, a high-quality end-to-end de novo assembler designed for extreme scale analysis, via efficient parallelization of the Meraculous code. Genome assembly software has many components, each of which stresses different components of a computer system. This chapter explains the computational challenges involved in each step of the HipMer pipeline, the key distributed data structures, and communication costs in detail. We present performance results of assembling the human genome and the large hexaploid wheat genome on large supercomputers up to tens of thousands of cores.

%\part{Extreme-Scale De Novo Genome Assembly}
\section{Overview of de novo Genome Assembly}

Genomes are the fundamental biochemical elements underlying inheritance, represented by chemical sequences of the four DNA ``letters" A, C, G, and T.  Genomes encode the basic software of an organism, defining the proteins that each cell can make, and the regulatory information that determines the conditions under which each protein is produced, allowing different organs and tissues to establish their distinct identities and maintain the stable existence of multicellular organisms like us.   Sequences that differ by as little as one letter can cause the expression of proteins that are defective or are inappropriately expressed at the wrong time or place. These differences underlie many inherited diseases and disease susceptibility.

Each organism's genome is a specific sequence, ranging in length from a few million letters for typical bacterium to 3.2 billion letters for a human chromosome to over 20 billion letters for some plant genomes, including conifers and bread wheat.  Genomes differ between species, and even between individuals within species; for example, two healthy human genomes typically differ at more than three million positions, and each can contain over ten million letters that are absent in the other.  Genomes mutate between every generation and even within individuals as they grow, and some of those mutations can drive cells to proliferate and migrate inappropriately, leading to diseases such as cancer.  We do not yet know which sequence differences are important but hope to learn these rules by sequencing millions of healthy and sick people, and comparing their genomes. 

Determining a genome sequence ``de novo" (that is, without reference to a previously determined sequence for a species) is a challenging computational problem. Modern sequencing instruments can cost-effectively produce only short sequence fragments of 100-250 letters, read at random from a genome (so-called ``shotgun" sequencing).  A billion such short reads can be produced for around \$1,000, enough to redundantly sample the human genome thirty times in overlapping short fragments. The computational challenge of ``genome assembly" is then to reconstruct chromosome sequences from billions of overlapping short sequence fragments, bridging a six order of magnitude gap between the length of the individual raw sequence reads and a complete chromosome.

Reconstructing a long sequence from short substrings is in general an NP-hard problem, and must rely on heuristics and/or take advantage of specific features of genome sequences~\cite{Simpson2015,Myers1995}. Current genome assembly algorithms typically rely on single node, large memory (e.g.\ 1 TB) architectures, and can take a week to assemble a single human genome, or even several months for larger genomes like loblolly pine~\cite{loblolly_Zimin}.  These approaches clearly do not scale to the assembly of millions of human genomes.  While some distributed memory parallel algorithms have been developed, they do not scale to massive concurrencies as they exhibit algorithmic bottlenecks and the irregular access patterns that are inherent to these algorithms amplify the distributed memory parallelization overheads.

%To add more urgency to the problem, over the last few years increasing attention has been devoted to the microbiome, the collective of single-celled organisms that live in diverse environments, including the human gut and other cavities and surfaces on the body.  The human microbiome contains several times as many cells as are in the human body, and the microbes that live in and on us are increasingly linked to health~\cite{Althani2015}.  But most of these microbes are undescribed and uncultivated, and their role in health or disease unknown.  An increasingly important way of discovering and characterizing these unknown microbes is direct shotgun sequencing of DNA from an environment, such as the human gut, producing a ``metagenomes".  If the genomes of the constituent microbes can be assembled from the resulting set of short fragments, we can gain valuable information about the functionl and role of these organisms.  The number of publicly available metagenomes datasets is rapidly increasing where in the span of just two years (2011-2013), the number of metagenome datasets nearly quadroupled and the number of identified protein coding genes increased 119 fold~\cite{imgm_2013}, far exceeding Moore's law of computing hardware.  An added complexity of metagenomes assembly is that in a given microbiome some species are common and others rare; each organism contributes to the raw data in proportion to its prevalence.   Hundreds or thousands of species can be present.

The work presented in this chapter addresses the aforementioned challenges by developing parallel algorithms for de novo assembly with the ambition to scale to massive concurrencies. The result of this work is HipMer~\cite{hipmer}, an end-to-end high performance de novo assembler designed to scale to massive concurrencies. HipMer uses (i) high performance computing clusters or supercomputers for both memory size and speed, (ii) a global address space programming model via Unified Parallel C (UPC)~\cite{upc} to permit random accesses across the aggregate machine memory, and (iii) parallel graph algorithms and hash tables, optimized for the statistical characteristics of the assembly process to reduce communication costs and increase parallelism. Our work is based on the Meraculous~\cite{meraculous,Merac2} assembler, a state-of-the-art de novo assembler for short reads developed at the Joint Genome Institute. Meraculous is a hybrid assembler that combines aspects of de Bruijn-graph-based assembly with overlap-layout-consensus approaches and is ranked at or near the top in most metrics of the Assemblathon II competition~\cite{Assemblathon2}. The original Meraculous used a combination of serial, shared memory parallel, and distributed memory parallel code. The size and complexity of genomes that could be assembled with Meraculous was limited by both speed and memory size. Our goal was a fast, scalable parallel implementation that could use the combined memory of a large scale parallel machine and our work ~\cite{sc14,ipdps15,hipmer,georganas2016scalable} has covered all aspects of the single genome assembly pipeline.

The rest of this chapter is organized as follows. Sections~\ref{sec:pipeline},\ref{sec:pgas},\ref{sec:hash} describe the fundamental concepts that we build upon in this chapter and provide the necessary background. Section~\ref{sec:parallel} details the parallelization in the HipMer algorithms. Section~\ref{sec:performance} presents performance results of HipMer on large scale. Section~\ref{sec:challenges} highlights the main challenges for porting HipMer to manycore architectures. Section~\ref{sec:related} briefly overviews related works and finally Section~\ref{sec:conclusion} concludes this chapter.

\section{The Meraculous Assembly Pipeline}
\label{sec:pipeline}

\begin{figure}[tb]
\centering
\includegraphics[width=0.9\textwidth]{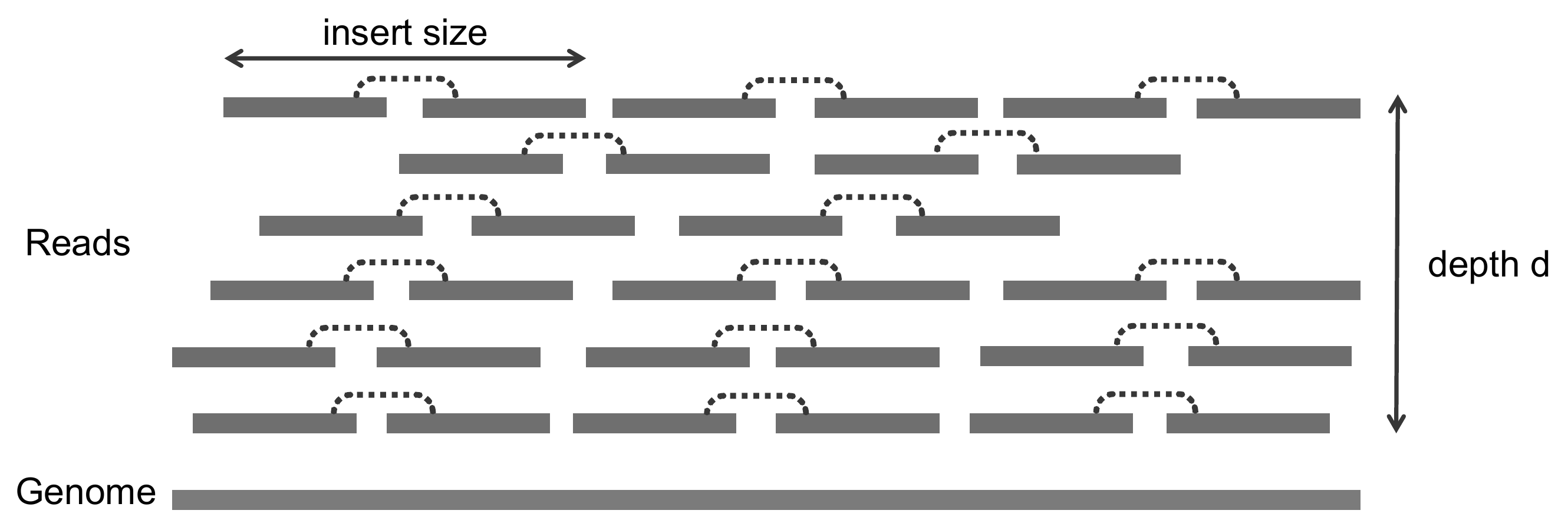}
\caption{Reads extracted from a genome with a depth of coverage $d$.}
\label{fig:reads}
\end{figure}
In this Section we review the Meraculous~\cite{meraculous,Merac2} single genome assembly pipeline and its main algorithmic components. This pipeline constitutes the basis for our parallel algorithms. Even though the description of the pipeline is specific to Meraculous, the high-level algorithmic techniques are relevant to any de novo genome assembler which is based on de Bruijn graphs.

The input to the genome assembler is a set of short, erroneous sequence fragments of 100-250 letters, read at random from a genome (see Figure~\ref{fig:reads}). Note that the genome is redundantly sampled at a depth of coverage $d$. Typically these reads fragments come in pairs and this information will be further exploited in the pipeline. Paired reads are also characterized by the \emph{insert size}, the distance between the two distant ends of the reads. Thus, given the read lengths and the corresponding insert size, we have an estimate for the gap between the paired reads. Typically the reads are grouped into libraries and each library is characterized by a nominal insert size and its standard deviation. Libraries with different insert sizes play a significant role in the assembly process, as will be explained later in this section. The Meraculous pipeline consists of four major stages (see Figure~\ref{fig:combined}(a)):

%\begin{figure}[tb]
%\centering
%\includegraphics[width=0.7\textwidth]{my_flow.pdf}
%\caption{Meraculous assembly flow chart.}
%\label{fig:assembly_flow}
%\end{figure}

\begin{figure}[tb]
\centering
\includegraphics[width=\textwidth]{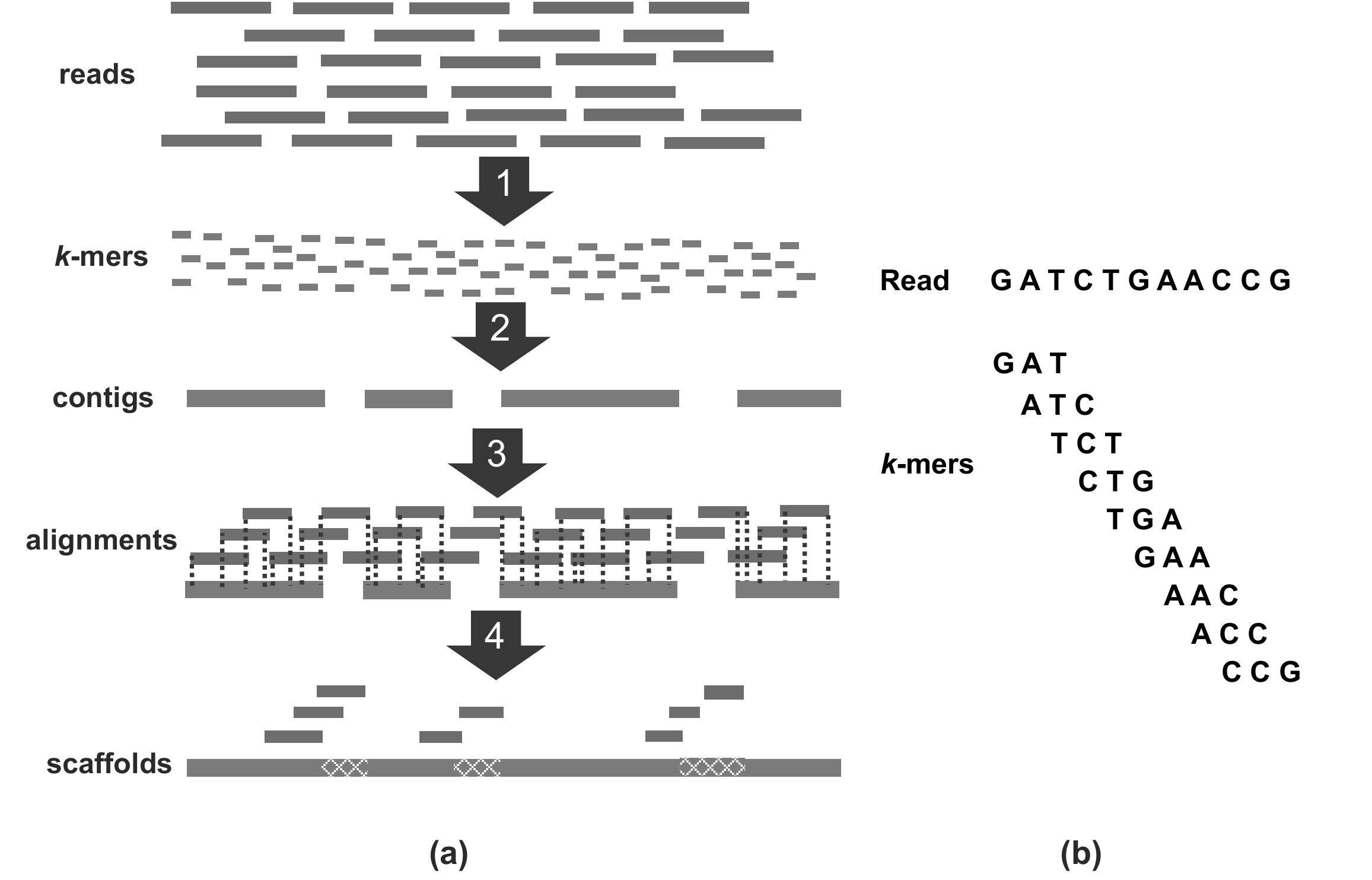}
\caption{(a) The Meraculous assembly pipeline. (b) Extracting $k$-mers ($k=3$) from the read \texttt{GATCTGAACCG}. }
\label{fig:combined}
\end{figure}

%\begin{figure}[tb]
%\centering
%\includegraphics[width=0.4\textwidth]{chopping.pdf}
%\caption{Extracting $k$-mers ($k=3$) from the read \texttt{GATCTGAACCG}.}
%\label{fig:chopping}
%\end{figure}

\textbf{1.\ K-mer analysis:} The input reads are processed to exclude errors. First, the reads are chopped into \emph{$k$-mers}, which are overlapping sequences of length $k$. Figure~\ref{fig:combined}(b) shows the $k$-mers (with $k=3$) that are extracted from a read. Then, the $k$-mers extracted from all the reads are counted and those that appear fewer times than a threshold are treated as erroneous and discarded. Additionally, for each $k$-mer we keep track of the two neighboring bases in the original read it was extracted from (henceforth we call these bases \emph{extensions}). The result of $k$-mer analysis is a set of $k$-mers and their corresponding extensions that with high probability include no errors.

The redundancy $d$ in the read data set is crucial in the process of excluding the errors implicitly. More specifically, an error at a specific read location yields up to $k$ erroneous $k$-mers. However, there are more reads covering the same genome location due to the redundancy $d$. More precisely, given the read length $L$ we expect to find a true, error-free $k$-mer on average $f =d \cdot (1-(k-1)/L)$ times in the read data set where $f$ is the mean of the Poisson distribution of key frequencies~\cite{liu2013estimation} and most of these $k$-mer occurrences will be error-free. Therefore, if we find a particular $k$-mer just one or two times in our read dataset, then we consider that to be erroneous. On the other hand, $k$-mers that appear a number of times proportional to $d$ are likely error-free.

\begin{figure}[tb]
\centering
\includegraphics[width=0.7\textwidth]{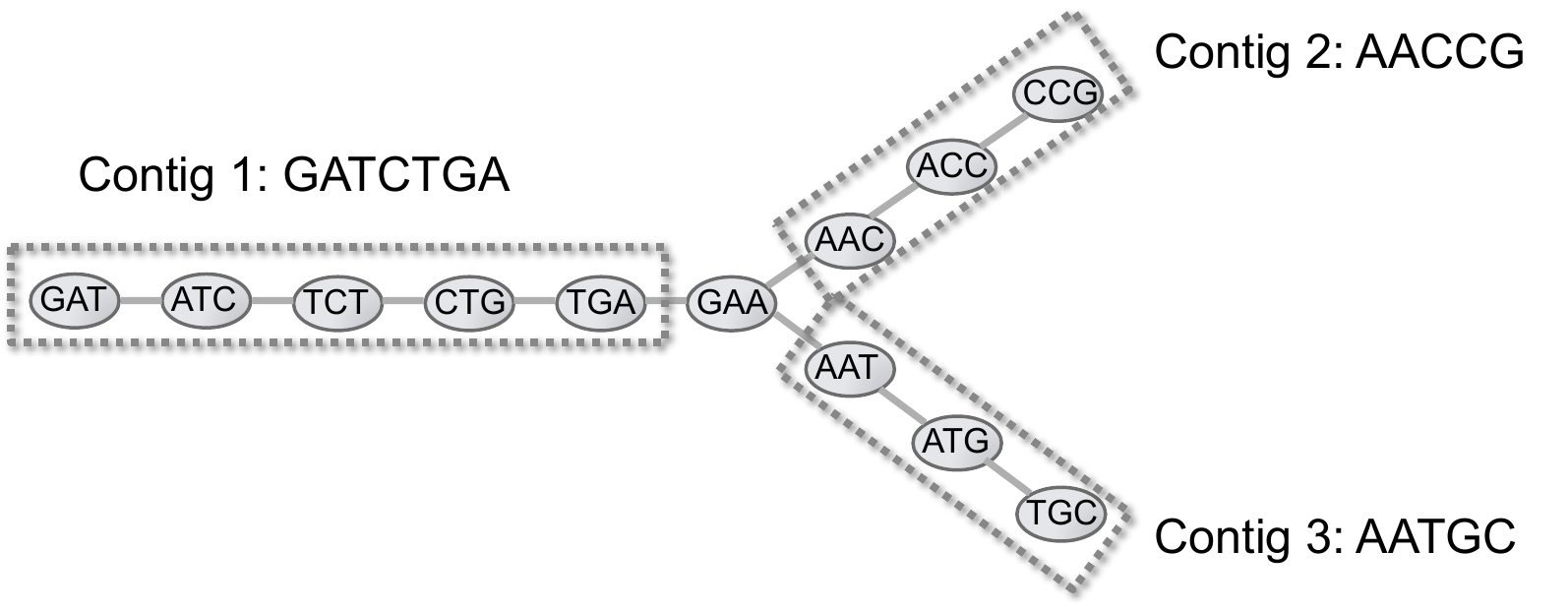}
\caption{A de Bruijn graph of $k$-mers with $k=3$.}
\label{fig:dbgraph1}
\end{figure}

\textbf{2.\ Contig generation:} The resulting $k$-mers from the previous step are stored in a de Bruijn graph. This is a special type of graph that represents overlaps in sequences. In this context, $k$-mers are the vertices in the graph, and two $k$-mers that overlap by $k-1$ consecutive bases are connected with an undirected edge in the graph (see Figure~\ref{fig:dbgraph1} for a de Bruijn graph example with $k= 3$). 

Due to the nature of DNA, the de Bruijn graph is extremely sparse. For example, the human genome's adjacency matrix that represents the de Bruijn graph is a $3\cdot 10^9 \times 3\cdot 10^9 $ matrix with between two and eight non-zeros per row for each of the possible extensions. In Meraculous only $k$-mers which have unique extensions in both directions are considered, thus each row has exactly two non-zeros.

Using a direct index for the $k$-mers is not practical for realistic values of $k$, since there are $4^k$ different $k$-mers. A compact representation can be leveraged via a hash table: A vertex ($k$-mer) is a key in the hash table and the incident vertices are stored implicitly as a two-letter code [\texttt{ACGT}][\texttt{ACGT}] that indicates the unique bases that immediately precede and follow the $k$-mer in the read dataset. By combining the key and the two-letter code, the neighboring vertices in the graph can be identified. 

In Figure~\ref{fig:dbgraph1} all $k$-mers (vertices) have unique extensions (neighbors) except from the vertex \texttt{GAA} that has two ``forward neighbors", vertices \texttt{AAC} and \texttt{AAT}. From the previous $k$-mer analysis results we can identify the vertices that do not have unique neighbors. In the contig generation step we exclude from the graph all the vertices with non-unique neighbors. We define \emph{contigs} as the connected components in the de Bruijn graph. Via construction and traversal of the underlying de Bruijn graph of $k$-mers the connected components in the graph are identified. The connected components have linear structure since we exclude from the graph all the ``fork" nodes or equivalently the $k$-mers with non-unique neighbors. The contigs are (with high probability) error-free sequences that are typically longer than the original reads. In Figure~\ref{fig:dbgraph1} by excluding the vertex \texttt{GAA} that doesn't have a unique neighbor in the ``forward" direction, we find three linear connected components that correspond to three contigs.

\begin{figure}[tb]
\centering
\includegraphics[width=\textwidth]{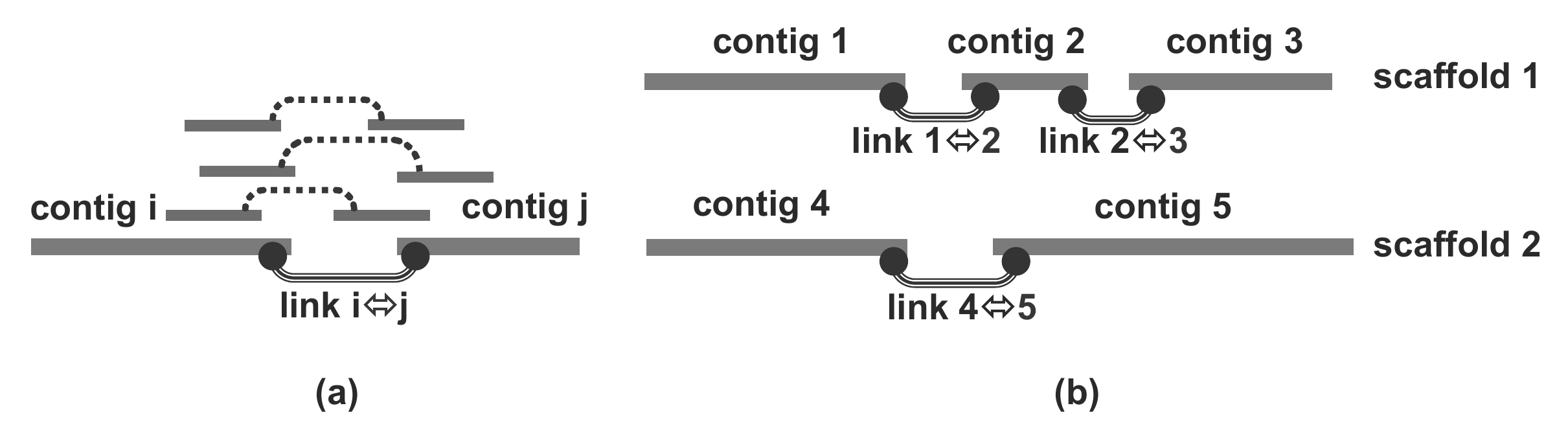}
\caption{(a) A link between contigs i and j that is supported by three read pairs. (b) Two scaffolds formed by traversing a graph of contigs.}
\label{fig:scaffolds}
\end{figure}

\textbf{3.\ Aligning reads onto contigs:} In this step we map the original reads onto the generated contigs. This mapping provides information about the relative ordering and orientation of the contigs and will be used in the final step of the assembly pipeline.

The Meraculous pipeline adopts a seed-and-extend algorithm in order to map the reads onto the contigs. First, the contig sequences are indexed by constructing a seed index, where the seeds are all substrings of length $k$ that are extracted from the contigs. This seed index is then used to locate candidate read-to-contig alignments. Given a read, we extract seeds of length $k$, look them up in the seed index and as a result we get candidate contigs that are aligned with the read because they share common seeds. Finally, an extension algorithm (e.g.\ Smith-Waterman~\cite{SW}) is applied to extend each found seed and local alignments are returned as the final result.

\textbf{4.\ Scaffolding and gap closing:} The scaffolding step aims to ``stitch" together contigs and form sequences of contigs called \emph{scaffolds} by assessing the paired-end information from the reads and the reads-to-contigs alignments. Figure~\ref{fig:scaffolds}(a) shows three pairs of reads that map onto the same pair of contigs i and j. Hence, we can generate a link that connects contigs i and j. By generating links for all the contigs that are supported by pairs of reads we create a graph of contigs (see Figure~\ref{fig:scaffolds}(b)). By traversing this graph of contigs we can form chains of contigs which constitute the scaffolds. Note that libraries with large insert sizes can be used to generate long-range links among contigs. Additionally, scaffolding can be performed in an iterative way by using links generated from different libraries at each iteration.

\begin{figure}[tb]
\centering
\includegraphics[width=0.6\textwidth]{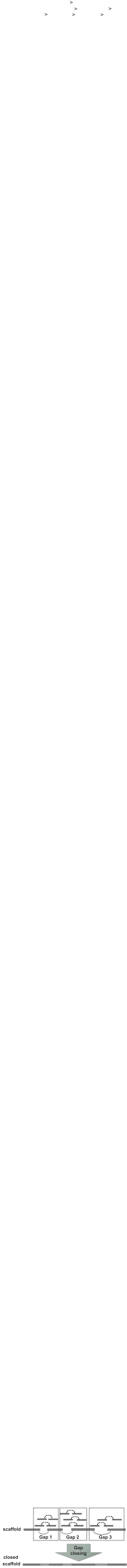}
\caption{The gap closing procedure.}
\label{fig:gapclosing}
\end{figure}

After the scaffold generation step, it is possible that there are gaps between pairs of contigs. We then further assess the reads-to-contigs mappings and locate the reads that are placed into these gaps (see Figure~\ref{fig:gapclosing}). Ultimately, we leverage this information and close the contig gaps by performing a mini-assembly algorithm involving only the localized reads for each gap. The outcome of this step constitutes the result of the Meraculous assembly pipeline.

%In Section~\ref{sec:parallel} we examine in more detail the algorithms involved in the Meraculous pipeline along with their parallelization.

In the subsequent Sections~\ref{sec:pgas},~\ref{sec:hash} and~\ref{sec:parallel} we will examine the programming model, the main distributed data structure and the parallel algorithms that are employed in the HipMer pipeline.

\section{The Partitioned Global Address Space Model in Unified Parallel C}
\label{sec:pgas}

The Partitioned Global Address Space (PGAS) programming model is employed in parallel programming languages. In this model, any thread is allowed to directly access memory on other threads. In the PGAS model, two threads may share the same physical address space or they may own distinct physical address spaces. In the former case, remote-thread accesses can be done directly using load and store instructions while in the latter case a remote access must be translated into a communication event, typically using a communication library such as GASNet~\cite{gasnet} or hardware specific layers such as Cray's DMAPP~\cite{dmapp} or IBM's PAMI~\cite{pami}.

An alternative communication mechanism typically employed in parallel programming languages is message passing, where the communication is done by exchanging messages between threads (e.g. see the Message Passing Interface (MPI)~\cite{mpi}). In such a communication model, both the sender and the receiver should explicitly participate in the communication event and therefore requires coordinating communication peers to avoid deadlocks. The programmer's burden in such a two-sided communication model can be further exaggerated in situations where the communication patterns are highly irregular as in distributed hash table construction. On the other hand, the PGAS model requires the explicit participation only of the peer that initiates the communication and as a result parallel programs with irregular accesses are easier to implement. Such a communication mechanism is typically referred to as \emph{one-sided communication}. In addition to PGAS languages like Unified Parallel C (UPC)~\cite{upc} there are programming libraries such as SHMEM~\cite{shmem} and MPI 3.0~\cite{mpi3} with one-sided communication features.

Unified Parallel C (UPC) is an extension of the C programming language designed for high performance computing on large-scale parallel machines by leveraging a PGAS communication model. UPC utilizes a Single Program Multiple Data (SPMD) model of computation in which the amount of parallelism is fixed at program startup time. On top of its one-sided communication capabilities, UPC provides global atomics, locks and collectives that facilitate the implementation of synchronization protocols and common communication patterns. In short, UPC combines the programmability advantages of the shared-memory programming paradigm and the control over data layout and performance of the message passing programming paradigm. According to the memory model of UPC each thread has a portion of shared and private address space. Variables that reside in the shared space can be directly accessed by any other thread and typically the program should employ synchronization mechanisms in order to avoid race conditions. On the other hand, variables that live in the private space can be read and written only by the thread owning that particular private address space.

Overall, the global address space model and the one-sided communication capabilities of UPC simplify the implementation of distributed data structures and highly irregular communication patterns. Such communication patterns are ubiquitous in our parallel algorithms described in Section~\ref{sec:parallel}.

%\begin{figure}[tb]
%\centering
%\includegraphics[width=0.7\textwidth]{pgas.pdf}
%\caption{An example of Partitioned Global Address Space in UPC.}
%\label{fig:pgas}
%\end{figure}

\section{Distributed Hash Tables in a PGAS Model}
\label{sec:hash}

A common data structure utilized in subsequent parallel algorithms is the distributed hash table. There is a wide body of work on concurrent hash tables~\cite{shun,hopscotch,hsu1986concurrent,ellis1987concurrency,kumar1990concurrent,michael2002high,shalev2006split} that focuses on shared memory architectures. There is also a lot of work on distributed hash tables (see~\cite{balakrishnan2003looking},~\cite{ratnasamy2001scalable} and survey of Zhang et al.~\cite{zhang2013survey}) specially designed for large-scale distributed environments that support primitive \texttt{put} and \texttt{get} operations. Such implementations do not target dedicated HPC environments and therefore have to deal with faults, malicious participants and system instabilities. Such distributed hash tables are optimized for execution on data centers rather than HPC systems with low-latency and high-throughput interconnects. There are some simple distributed memory implementations of hash tables in MPI~\cite{gerstenberger2013enabling} and UPC~\cite{maynard2012comparing}, but they are used mainly for benchmarking purposes of the underlying runtime and do not optimize the various operations depending on the use case of the hash table. In this section we describe the basic implementation of a distributed hash table using a PGAS abstraction. We also identify a handful of use cases for distributed hash tables that enable numerous optimizations for HPC environments.

%In this section we describe the basic implementation of a distributed hash table in a PGAS model since this is a common distributed data structure utilized in subsequent parallel algorithms. We also identify a handful of use cases for distributed hash tables that enable numerous optimizations.  

\subsection{Basic implementation of a distributed hash table}

\begin{figure}[tb]
\centering
\includegraphics[width=\textwidth]{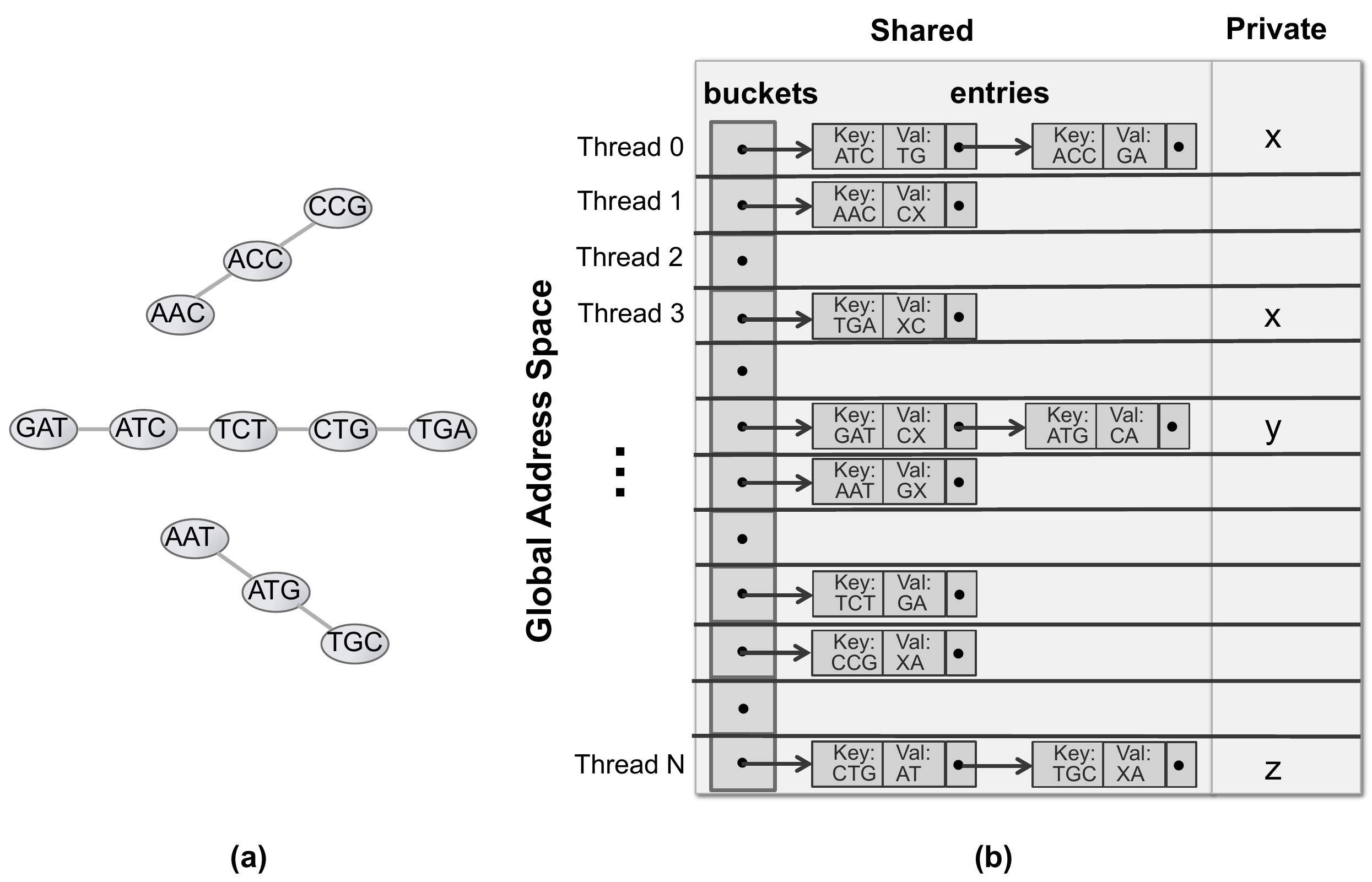}
\caption{(a) A de Bruijn graph of $k$-mers ($k=3$). (b) A distributed hash table that represents de Bruijn graph at left.}
\label{fig:dbghash}
\end{figure}

We will present the vanilla implementation of a distributed hash table by following an example of a distributed de Bruijn graph. Figure~\ref{fig:dbghash} (a) shows a de Bruijn graph of $k$-mers with $k=3$ and Figure~\ref{fig:dbghash} (b) illustrates its representation in a distributed hash table.  A vertex ($k$-mer) in the graph is a key in the hash table and the incident vertices are stored implicitly as a two-letter code [\texttt{ACGTX}][\texttt{ACGTX}] that indicates the unique bases that follow and  precede that $k$-mer. This two letter code is the value member in a hash table entry. Note that the character \texttt{X} indicates that there is no neighboring vertex in that direction. By combining the key and the two-letter code, the neighboring vertices in the graph can be identified. More specifically, by concatenating the last $k-1$ letters of a key and the first letter of the value, we get the ``forward" neighboring vertex. Similarly, by concatenating the second letter of the value and the first $k-1$ letters of that key, we get the ``preceding" neighboring vertex.

In our example, all the hash table entries are stored in the shared address space and thus they can be accessed by any thread. The buckets are distributed to the available threads in a cyclic fashion to achieve load balance. Our hash table implementation utilizes a chaining rule to resolve collisions in the buckets (entries with the same hash value). We emphasize here that the hash tables involved in our algorithms can be gigantic (hundreds of Gbytes up to tens of Tbytes) and cannot fit in a typical shared-memory node. Therefore it is crucial to distribute the hash table buckets over multiple nodes and in this quest the global address space of UPC is convenient.

In the following subsection we list different use case scenarios of distributed hash tables. These use case scenarios are encountered in our parallel algorithms described in Section~\ref{sec:parallel} and are presented upfront in order to highlight the optimization opportunities.

\subsection{Use cases of distributed hash tables in the HipMer pipeline}
\label{subsec:usecases}

Here we identify a handful of use cases for the distributed hash tables that allow specific optimizations in their implementation. These use cases will be used as points of reference in the section that details our parallel algorithms.

\textbf{Use case 1 -- Global Update-Only phase (GUO)}:
The operations performed in the distributed hash table are only global updates with commutative properties (e.g.\ inserts only). The global hash table will have the same state regardless of insert order, although it might possibly have different underlying representation due to chaining. The global update-only phase can be optimized by dynamically aggregating fine-grained updates (e.g.\ inserts) into batch updates. In this way we can reduce the number of messages and synchronization events. We can also overlap computation/communication or pipeline communication events to further hide the communication overhead.

A typical example of such a use case is a producer/consumer setting where the producers operate in a distinct phase from consumers, e.g.\ all consumers insert items in a hash table before anything is consumed/read.

\textbf{Use case 2 -- Global Reads \& Writes phase (GRW)}:
The operations performed during this phase are global reads and writes over the \emph{already inserted entries}. Typically we can't batch reads and/or writes since there might be race conditions that affect the control flow of the governing parallel algorithm. However, we can use global atomics (e.g.\ compare-and-swap) instead of fine-grained locking in order to ensure atomicity. The global atomics might employ hardware support depending on the platform and the corresponding runtime implementation. We can also build synchronization protocols at a higher level that do not involve the hash table directly but instead are triggered by the results of the atomic operations. Finally, we can implement the delete operation of entries with atomics and avoid locking schemes.

For example, consider the consumers in a producer/consumer scenario that compete for the entries of the hash table. The entries may have utilization signatures (i.e.\ ``used" binary flags) that can be accessed via global atomics and indicate whether the corresponding entries have been consumed or not. An orthogonal optimization for this use-case scenario is to adopt locality sensitive hashing schemes to increase locality and decrease communication volume/latency overhead of global atomics.

\textbf{Use case 3 -- Global Read-Only phase (GRO)}:
In such a use case, the entries of the distributed hash table are read-only and a degree of data reuse is expected. The optimization that can be readily employed is to design software caching schemes to take advantage of data reuse and minimize communication. These caching frameworks can be viewed as ``on demand" copying of remote parts of the hash table. Note that the read-only phase guarantees that we do not need to provision for consistency across the software caches. Such caching optimizations can be used in conjunction with locality-aware partitioning to increase effectiveness of the expected data reuse. Initially even if the data is remote, it is likely to be reused later locally.

A typical example of this use case is a lookup-only hash table that implements a database/index. This is a special case of the consumer side in a producer/ consumer setting where the entries can be consumed an infinite number of times. 

\textbf{Use case 4 -- Local Reads \& Writes phase (LRW)}:
In this use case, the entries in the hash table will be further read/written only by the processor owning them. The optimization strategy we employ in such a setting is to use a deterministic hashing from the sender side and local hash tables on the receiver side. The local hash tables ensure that we avoid runtime overheads. Additionally, high-performance, serial hash table implementations can be seamlessly incorporated into parallel algorithms.

For example, consider items that are initially scattered throughout the processors and we want to send occurrences of the same item to a particular processor for further processing (e.g.\ consider a ``word-count" type of task). Each processor can insert the received items into a local hash table and further read/write the local entries from there.

\begin{table}[t!]
\centering
\begin{tabular}{cccccc}
  Use case & \rot[20][1.6cm]{\shortstack[l]{Dynamic message\\aggregation}}
  & \rot[20][1.5cm]{\shortstack[l]{Remote global\\atomics}}
  & \rot[20][1.5cm]{\shortstack[l]{Caching of remote\\entries}}
  & \rot[20][1.5cm]{\shortstack[l]{Locality sensitive\\hashing}}
  & \rot[20][1.5cm]{\shortstack[l]{Serial hash\\table library}} \\
  \midrule
  GUO   &  \checkmark & \checkmark  &  & \checkmark &  \\ \hline
  GRW  &  & \checkmark &  & \checkmark &\\ \hline
  GRO  &  &  & \checkmark & \checkmark &\\ \hline
  LRW  & \checkmark &  &  & \checkmark & \checkmark \\
\end{tabular}
\caption{Distributed hash table optimizations for various use case scenarios.}
\label{tab:opts}
\end{table}

We emphasize that this is not an exhaustive list of use cases for distributed hash tables. Nevertheless, it captures the majority of the computational patterns we identified in our parallel algorithms that will be detailed in the following Section~\ref{sec:parallel}. Table~\ref{tab:opts} summarizes  the optimizations we can employ for the various use cases of the distributed hash tables. Multiple of the aforementioned use cases can be encountered during the lifetime of a distributed hash table; in most of the cases the optimizations can be easily composed (e.g.\ by having semantic barriers to signal the temporal boundaries of the phases). For example, the Global Update-Only phase can be followed by a Global Read-Only phase in a scenario where a database is first built via insertion of the corresponding items into a hash table and later the distributed data structure is reused as a global lookup table.

\section{Parallel algorithms in HipMer}
\label{sec:parallel}
In this Section we detail the parallelization of the Meraculous pipeline presented in Section~\ref{sec:pipeline}. In our description we refer to ideas from Sections~\ref{sec:pgas} and~\ref{sec:hash} regarding the PGAS programming model and distributed hash tables.

\subsection{Parallel $k$-mer analysis}
Counting the frequencies of each distinct $k$-mer involves 
reading the input DNA short reads, parsing the reads into
$k$-mers, and keeping a count of each distinct $k$-mer 
that occurs more than $\epsilon$ times ($\epsilon \approx 1, 2$). 
The reason for such a cutoff is to eliminate sequencing errors.
$K$-mer analysis additionally requires keeping track of all possible extensions of the $k$-mer from either side.
This is performed by keeping two short integer arrays of length four per $k$-mer, where each entry in the array keeps track
of the number of occurrences of each nucleotide [\texttt{ACGT}] on either end. If a nucleotide on an end appears more times than a threshold $t_{hq}$, it is characterized as high quality extension. One of the difficulties with performing $k$-mer analysis in
distributed memory is that the size of the intermediate data (the set
of $k$-mers) is significantly larger than both the input and the output, since each read is subsequenced with overlaps of $k-1$ base pairs. 

As each processor reads a portion of the reads and extracts the corresponding $k$-mers, a deterministic map function maps each $k$-mer to a processor id. Once the $k$-mers are generated, we perform an irregular all-to-all exchange step in order to communicate the $k$-mers among the processors based on the calculated processor ids. This deterministic mapping assigns all the occurrences of a particular sequence to the same processor, thus eliminating the need for a global hash table; instead, each processor maintains a local hash table to count the occurrences of the received $k$-mers. We refer to this model of computation as ``owner-computes". Note that this computational pattern fits the Use Case 4 (LRW) of the distributed hash tables. Given the genome size $G$, the coverage $d$ and the read length $L$, the total number of $k$-mers that have to be communicated are $\Theta(\frac{Gd}{L}(L-k+1))$.

In this parallel algorithm, memory consumption quickly becomes a problem due to errors because a single nucleotide error creates up to $k$ erroneous $k$-mers.
It is not uncommon to have over $80$\% of all distinct $k$-mers erroneous, depending on the read length and the value of $k$.
We ameliorate this problem using Bloom filters, which were previously used in serial $k$-mer 
counters~\cite{melsted2011efficient}.  A Bloom filter~\cite{bloom1970space} is a space-efficient probabilistic data structure used for membership queries. It might have false positives, but no false negatives. If a $k$-mer 
was not seen before, the filter can accidentally report it as `seen'. However, if a $k$-mer 
was previously inserted, the Bloom filter will certainly report it as `seen'. This is suitable for $k$-mer counting as no real $k$-mers will be missed. If the Bloom filter reports that a $k$-mer was seen before, then the corresponding processor inserts that $k$-mer to the actual local hash table in order to perform the counting.
Our novelty is the discovery that localization of $k$-mers via the deterministic $k$-mer to processor id mapping is \emph{necessary and sufficient}
to extend the benefits of Bloom filters to distributed memory. 

The false positive rate of a Bloom filter is 
$Pr(e) = (1-e^{-hn/m})^h$ for $m$ being the number of distinct elements in the dataset, $n$ the size of the Bloom filter, and $h$ the number of hash functions used. There is an optimal number of hash functions given $n$ and $m$, which is $h = \ln 2  \cdot (m / n)$. In practice, we achieve approximately $5\%$ false positive rate using only $1$-$2$\% of the memory that would be needed to 
store the data directly in a hash table (without the Bloom filter). Hence, in a typical dataset where $80\%$ of all $k$-mers are errors, 
we are able to filter out $76\%$ of all 
the $k$-mers using almost no additional memory. Hence, we can
effectively run a given problem size on a quarter of the nodes 
that would otherwise be required.

We have so far ignored that Bloom filters need to know the number of distinct elements expected to perform optimally. While
dynamically resizing a Bloom filter is possible, it is expensive
to do so. We therefore use a cardinality estimation algorithm to approximate the number of 
distinct $k$-mers. Specifically, we use the Hyperloglog algorithm~\cite{DMTCSPROCdmAH0110},
 which achieves less than $1.04/\sqrt{m}$ error for a dataset of $m$ distinct elements. Hyperloglog requires a
only several KBs of memory to count trillions of items. The basic idea behind cardinality estimators is hashing each 
item uniformly and maintaining the minimum hash value.
 Hyperloglog maintains multiple buckets for increased accuracy and uses the number of trailing zeros in the bitwise representation 
 of each bucket as the estimator. 
 
The observation that leads to minimal communication parallelization of Hyperloglog is as follows.
Merging Hyperloglog counts for multiple datasets can be done by keeping the minimum of their final buckets by a parallel reduction. Consequently, the communication volume for this first cardinality estimation pass 
is \emph{independent of the size of the sequence data}, and is only a function of the Hyperloglog data structure size. 
In practice, we implement a modified version of the algorithm that
uses 64-bit hash values as the original 32-bit hash described in the
original study~\cite{DMTCSPROCdmAH0110} is not able to process our massive datasets.

One downside of this parallel counting approach is that highly complex plant genomes, such as wheat, are extremely repetitive and it is not uncommon to see $k$-mers that occur millions of times. Such high-frequency $k$-mers create a significant load imbalance problem, as the processors assigned to these high-frequency $k$-mers require significantly more memory and processing times. We improve our approach by first identifying frequent $k$-mers (i.e.``heavy hitters'' in database literature) and treating them specially~\cite{hipmer}. In particular, the ``owner-computes" method is still used for low-to-medium frequency $k$-mers but the high frequency $k$-mers are accumulated locally on each processor, followed by a final global reduction. Since an initial pass over the data is already performed to estimate the cardinality (the number of distinct $k$-mers) and efficiently initialize our Bloom filters, running a streaming algorithm for identifying frequent $k$-mers during the same pass is essentially free.

%In our work use the counter-based algorithm of Misra and Gries~\cite{misra1982finding} (subsequently reinvented several times~\cite{Demaine:2002:FEI, karp2003simple}). 
\subsection{Parallel contig generation}
\label{subsec:contig}
Once we have performed the $k$-mer analysis step, it is necessary to store the resulting $k$-mers in a distributed hash table that represent the de Bruijn graph in a compact way. A vertex ($k$-mer) is a key in the hash table and the incident vertices are stored implicitly as a two-letter code [\texttt{ACGT}][\texttt{ACGT}] that indicates the unique bases that immediately precede and follow the $k$-mer in the read dataset. These graphs typically are huge and require hundreds of GBs or even TBs for large genomes in order to be stored in memory. Therefore, we employ the global address space of Unified Parallel C (UPC) in order to transparently store the distributed hash table in distributed memory and overcome the limitations of requiring specialized, large shared memory machines.

During the parallel hash table construction, we consider only the $k$-mers that have unique high-quality extensions in \emph{both directions}. These $k$-mers are hashed and sent to the proper (potentially remote) bucket of the hash table by leveraging the one-sided communication capabilities of UPC. We recognize this computational pattern as the Use Case 1 (GUO) of the distributed hash tables, therefore we can mitigate the communication and synchronization overheads by leveraging dynamic message aggregation. In particular, we designed a dynamic aggregation algorithm~\cite{sc14} where the $k$-mers are aggregated in batches before being sent to the appropriate processors. The pattern deployed here is also an {irregular all-to-all communication}. However, unlike $k$-mer analysis, the total number of $k$-mers that have to be communicated is $\Theta(G)$, since multiple occurrences of $k$-mers have been collapsed during the $k$-mer analysis stage and this condensed $k$-mer set should have size proportional to the genome size $G$. 

Once the distributed de Bruijn graph (hash table) has been constructed, we traverse it in parallel and identify the connected components that represent the \emph{contig} sequences. Typically such de Bruijn graphs have extremely high-diameter (the connected components in theory can have size up to the length of chromosomes) and therefore traditional parallelization strategies of the graph traversal would not scale to extreme concurrencies.

\begin{figure}[!tb]
\centering
\includegraphics[width=\textwidth]{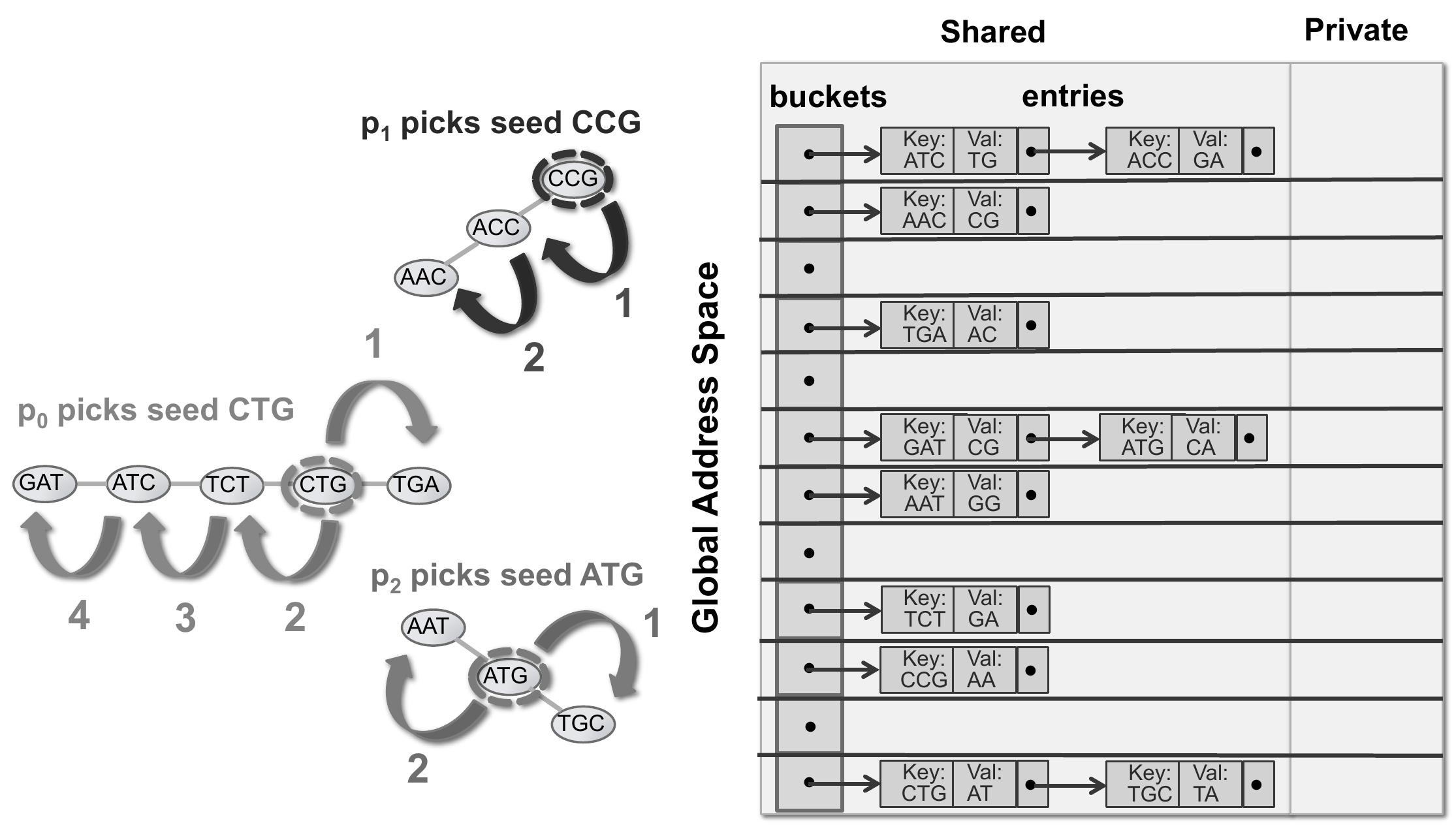}
\caption{Parallel de Bruijn graph traversal. Processor 0 picks a $k$-mer called ``traversal seed" (vertex \texttt{CTG}) and with four lookups in the distributed hash tables it explores the four remaining vertices of that connected component. The numbered arrows indicate the order in which processor 0 looks up the corresponding vertices in the distributed hash table. In an analogous way, processors 1 and 2 pick seeds \texttt{CCG} and \texttt{ATG} respectively and explore in parallel with processor 0 different connected components of the underlying de Bruijn graph.}
\label{fig:graph_traversal}
\end{figure}

In order to form a contig, a processor $p_i$ chooses a random $k$-mer from its own part of the distributed hash table as seed and creates a new \emph{subcontig} (incomplete contig) data structure which is represented as a string and the initial content of the string is the seed $k$-mer. Processor $p_i$ then attempts to extend the subcontig towards both of its endpoints using the high quality extensions stored as values in the distributed hash table. To extend a subcontig  from its right endpoint, processor $p_i$ uses the $k-1$ last bases and the right high quality extension $R$ from the right-most $k$-mer in the subcontig. 
It therefore concatenates the last $k-1$ bases and the extension $R$ to form the next $k$-mer to be searched in the hash table. Processor $p_i$ performs a lookup for the newly formed $k$-mer and if it is found successfully, the subcontig is extended to the right by the base $R$. The same process can be repeated until the lookup in the hash table fails, meaning that there are no more \texttt{UU} $k$-mers that could extend this subcontig in the right direction. A subcontig can be extended to its left endpoint using an analogous procedure. If processor $p_i$ can not add more bases to either endpoint of the subcontig, then a contig has been formed (or equivalently a connected component in the de Bruijn graph has been explored) and is stored accordingly.

Figure~\ref{fig:graph_traversal} illustrates how the parallel algorithm works with three processors. Processor 0 picks a random traversal seed (vertex \texttt{CTG}) and initializes a subcontig with content \texttt{CTG}. Then, by looking in the distributed hash table the entry \texttt{CTG} it gets back the value \texttt{AT}, meaning that the right extension is \texttt{A} and the left extension is \texttt{T}. After that, processor 0 forms the next $k$-mer to be looked up (\texttt{TGA}) by concatenating the last 2 bases of \texttt{CTG} and the right extension \texttt{A} -- this lookup corresponds to the arrow 1 of processor 0.  By following the analogous procedure and three more lookups in the distributed hash table, processor 0 explores all the vertices of that connected component that corresponds to the contig \texttt{GATCTGA}. The numbered arrows indicate the order in which processor 0 looks up the corresponding vertices in the distributed hash table. In an analogous way, processors 1 and 2 pick seeds \texttt{CCG} and \texttt{ATG} respectively and explore in parallel with processor 0 different connected components of the underlying de Bruijn graph.

All processors independently start building subcontigs and no synchronization is required unless two processors pick initial $k$-mer seeds that eventually belong in the same contig. In this case, the processors have to collaborate and resolve this conflict in order to avoid redundant work and race conditions. The high-level idea of the synchronization protocol for conflict resolution is that one of the involved processors backs off, and the other processor takes over the computed ``subcontig" from the processor that backed off. We designed a lightweight synchronization scheme~\cite{sc14} based on {remote atomics} and in our previous work we proved (under some modeling assumptions) that our synchronization algorithm is scalable to massive concurrencies. Finally, the parallel traversal is terminated when all the connected components in the de Bruijn graph are explored.

The access pattern in the distributed hash table consists of highly {irregular, fine-grained lookup operations}. The size of the de Bruijn graph is proportional to the genome size, thus the traversal involves visiting $\Theta(G)$ vertices via atomics and irregular lookup operations. The computational task of the graph traversal is to visit all the already inserted $k$-mers in the distributed hash table. During this parallel procedure, we cannot batch reads and/or writes since there might be race conditions that affect the control flow of the synchronization algorithm. However, we use global atomics instead of fine-grained locking and we build synchronization protocols at a higher level that do not involve the distributed hash table directly but instead are triggered by the results of the atomic operations on the objects stored inside the hash table. We recognize this computational pattern as the Use Case 2 (GRW) of the distributed hash tables.

\subsection{Parallel read-to-contig alignment}

Here we describe the parallel algorithm that maps the original reads onto the contig sequences. First, each processor reads a distinct portion of the contig
sequences and stores them in global address space  such that any other processor can access them. Every contig sequence of length $C$ contains $C-k+1$ distinct seeds (substrings) of length $k$. We extract in parallel seeds from the contigs and associate with every seed the contig from which it was extracted. Since the contigs constitute a fragmented version of the genome, in total we extract $\Theta(G)$ seeds.

\begin{figure}[!t]
\centering
\includegraphics[width=0.85\textwidth]{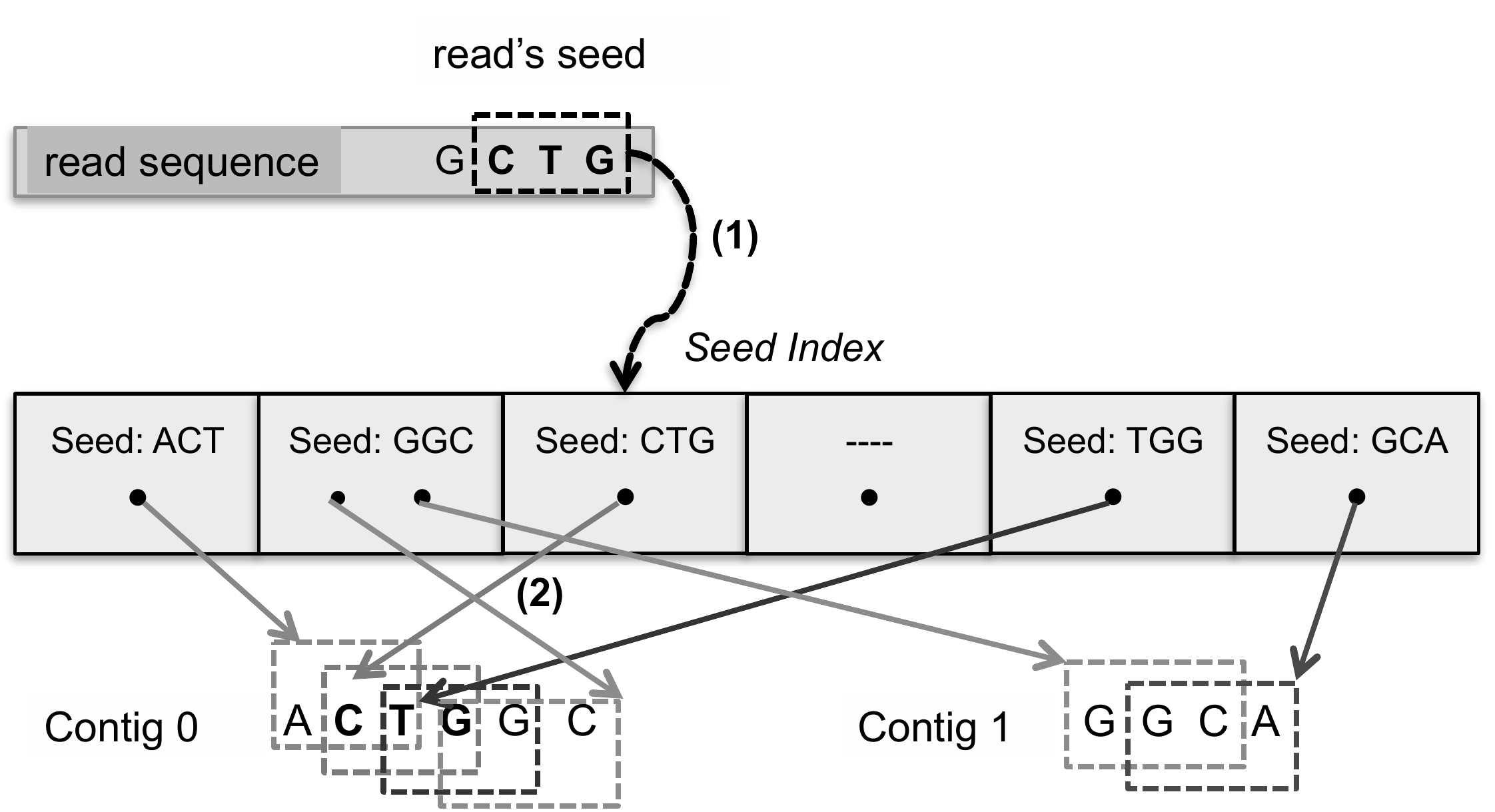}
\caption{Locating read-to-contig candidate alignments. First the
  processor extracts a seed from the read sequence (\texttt{CTG}
  seed). Next,  the processor looks up the distributed seed index (arrow 1) and finds that a candidate contig sequence is Contig 0 (arrow 2). Finally, the Smith-Waterman algorithm is executed using as inputs the read and the Contig 0 sequences.}
\label{fig:lookup}
\end{figure}

Once the seeds are extracted from the contigs, they are stored in a global hash table, referred to as the seed index where the key is a seed and the value is a pointer to the contig from which this seed has been extracted. The seed index is distributed and stored in global shared memory such that any processor can access and lookup any seed. Essentially the seed index data structure provides a mapping from seeds to contigs. The seeds are stored in the global seed index via an irregular {all-to-all communication} step similar to the hash table construction in the contig generation phase. Again, we recognize this computational pattern as the Use Case 1 (GUO) of the distributed hash tables, therefore we can mitigate the communication and synchronization overheads by leveraging dynamic message aggregation. Figure~\ref{fig:lookup} illustrates how two contigs are indexed by using seeds with length $k=3$.

After the seed index construction, we proceed with the aligning phase where every read is mapped onto contigs. Initially, each processor is assigned an equal number of reads. For each read of length $L$, a processor extracts all $L-k+1$ seeds of length $k$ contained in it. Given a seed $s$ from a read, the processor performs a {fine-grained lookup} in the global seed index and locates the candidate contigs that have in common the seed $s$ with that read. Thus, each one of the read-to-contig candidate alignments can be located in $\Theta(1)$ time. Figure~\ref{fig:lookup} exhibits an example of how we can locate a read-to-contig candidate alignment by leveraging the seed index. We emphasize that in the alignment phase, all processors operate concurrently on distinct reads.

 If we naively execute an exhaustive lookup of all possible seeds, in total we have to perform $O(\frac{Gd}{L}(L-k+1))$ lookups. Our optimized parallel algorithm though~\cite{ipdps15} identifies properties in the contigs during the hash index construction that reduce significantly the number of lookups.

We also made the observation that our parallel alignment phase makes no writes/updates in the distributed seed index or the distributed data structure that stores the contig sequences after their construction phase; it just uses them for lookups/reads. We recognize this computational pattern as the Use Case 3 (GRO) of the distributed hash tables and our parallel algorithm~\cite{ipdps15} exploits software caches to maximize data reuse and avoids off-node lookups.

Finally, after locating a candidate contig that has a matching seed with the read we are processing, the Smith-Waterman algorithm is executed with input the read and contig sequences in order to perform local sequence alignment. The output of this stage is a set of reads-to-contig alignments.

\subsection{Parallel scaffolding and gap closing}
The first part of scaffolding involves processing of the reads-to-contig alignments (Figure~\ref{fig:scaffolds}(a)) and generating links among contigs. In order to parallelize this operation, we index only the relevant alignments (those that indicate that two contigs should be connected) via a distributed hash table. This distributed hash table construction employs an {irregular all-to-all communication} pattern similar to the contig generation stage (Use Case 1 of distributed hash tables). We emphasize that the graph of contigs (and consequently the number of links among them) is orders of magnitude smaller that the $k$-mer de Bruijn graph because the connected components in the $k$-mer graph are now contracted to single vertices in the contig-graph. According to the Lander-Waterman statistics~\cite{expected}, the expected number of contigs is $\Theta(dG/L\cdot e^{-d})$.

Then, we process the contigs to identify properties (e.g.\ average $k$-mer depth, termination states) that will help us further simplify the contig-graph. This step necessitates looking up $k$-mer info in a global hash table of $k$-mers with $\Theta(G)$ size. Afterwards, we introspect the contig-graph to identify bubble structures via a parallel traversal which requires {irregular lookups} in the distributed contig-graph representation and {global atomics} (Use Case 2 of the distributed hash tables). After the bubble removal step, we traverse the  simplified graph and generate scaffolds (Figure~\ref{fig:scaffolds}(b)). The last traversal is done by selecting starting vertices in order of decreasing length (this heuristic tries to stitch together first ``long" contigs) and therefore it is inherently serial. We have optimized this component and found that its execution time is insignificant compared to the previous pipeline operations -- this behavior is expected as the input contig-graph is relatively small as explained earlier.

The gap closing stage uses the read-to-contig alignments, the scaffolds and the contigs to attempt to assemble reads across gaps between the contigs of scaffolds (see Figure~\ref{fig:scaffolds}(b)). To determine which reads map to which gaps, the alignments are processed in parallel and projected into the gaps. We utilize a distributed hash table to localize the unassembled reads onto the appropriate gaps via {irregular all-to-all communication}. Assuming that a fraction $\gamma$ of the genome is not assembled into contigs, then this communication step involves $\Theta(\gamma Gd/L)$ reads. Finally, the gaps are divided into subsets and each set is processed by a separate processor, in an embarrassingly parallel phase.

%Table~\ref{tab:patterns} summarizes the communication operations in the parallel algorithms. Each stage involves though different volume of data and, as described earlier, these data are communicated with different communication patterns.

\subsection{Summary of communication patterns and costs}
\label{subsec:costs}
Given the genome size $G$, the read length $L$, the coverage $d$, and the fraction $\gamma$ of the reads that are not assembled into contigs, Table~\ref{tab:volume} summarizes for each stage the main communication patterns along with the corresponding volume of communication. These communications patterns govern the efficiency of the parallel pipeline at large scale, where most of the stages are communication-bound. The different communication patterns have, however, vastly different overheads. For example, the all-to-all communication exchange is typically bounded by the bisection bandwidth of the system assuming that the partial messages are large enough and there is enough concurrency to saturate the available bandwidth. On the other hand, fine-grained, irregular lookups and global atomics are typically latency-bound and their efficiency relies upon the ability of the interconnect to serve those fine-grained, irregular request efficiently at high concurrencies.

%\begin{table}[t!]
%\centering
%\begin{tabular}{cccc}
% \textbf{Stage} & {{\shortstack[c]{All-to-all\\exchange}}} & {{\shortstack[c]{Irregular\\lookups}}}  & {{\shortstack[c]{Remote\\atomics}}}  \\ \hline
%  $k$-mer analysis  &  \checkmark & &  \\ \hline
%  Contig Generation & \checkmark  & \checkmark & \checkmark   \\ \hline
%  Sequence alignment & \checkmark & \checkmark  &  \checkmark  \\ \hline
%  Scaffolding \& Gap Closing  & \checkmark  &   \checkmark & \checkmark  \\ \hline
%\end{tabular}
%\caption{Communication operation used in the HipMer pipeline}
%\label{tab:patterns}
%\vspace{-10pt}
%\end{table}

\begin{table}[t!]
\centering
\begin{tabular}{cccc}
 \textbf{Stage} & {\textbf{Communication pattern}} & \textbf{Volume of data}    \\ \hline
  $k$-mer analysis  &  all-to-all exchange & $\Theta(Gd\cdot(L-k+1)/L)$  \\ \hline 
   	Contig		& all-to-all exchange  & $\Theta(G)$    \\ 
   Generation & irregular lookups &  $\Theta(G)$ \\
					& global atomics & $\Theta(G)$ \\  \hline
 Sequence  & all-to-all exchange & $\Theta(G)$   \\ 
  	alignment				    & irregular lookups & $O(Gd\cdot(L-k+1)/L)$   \\ \hline
   				& all-to-all exchange  & $\Theta(G)$    \\ 
Scaffolding  & irregular lookups & $\Theta(G)$ \\
					& global atomics & $\Theta(dG/L\cdot e^{-d})$ \\  \hline
Gap closing  &  all-to-all exchange & $\Theta(\gamma Gd/L)$  \\ 
\end{tabular}
\caption{Major communication operations in the HipMer pipeline}
\label{tab:volume}
\vspace{-10pt}
\end{table}
\section{Performance Results}
\label{sec:performance}

Parallel performance experiments are conducted on Edison, the Cray
XC30 located at NERSC. Edison has a peak performance of 2.57
petaflops/sec, with 5,576 compute nodes, each equipped with 64 GB RAM
and two 12-core 2.4GHz Intel Ivy Bridge processors for a total of
133,824 compute cores, and interconnected with the Cray Aries network
using a Dragonfly topology.  For our experiments, we use Edison's parallel Lustre \texttt{/scratch3} file system, which has 144 Object Storage Servers providing 144-way concurrent access to the I/O system with an aggregate peak performance of 72 GB/sec.

To analyze HipMer performance behavior we
examine a human genome for a member of
the CEU HapMap population (identifier NA12878) sequenced by the Broad
Institute. The genome contains 3.2 Gbp (billion base pair) 
assembled from 2.9 billion reads (290 Gbp of sequence), which are 101 bp in length, from a paired-end insert
library with mean insert size 395 bp. Additionally, we examine the
grand-challenge hexaploid wheat genome (Triticum aestivum L.)
containing 17 Gbp from 2.3 billion reads (477 Gbp of sequence) for the
homozygous bread wheat line `Synthetic W7984'.  Wheat reads are
150-250 bp in length from 5 paired-end libraries with insert sizes
240-740 bp. Also, for the scaffolding phase we leveraged (in addition to the previous libraries) two long-insert paired-end DNA libraries with insert sizes 1 kbp and 4.2 kbp. This important genome was only recently sequenced for the
first time~\cite{mayer2014chromosome}, and requires high-performance analysis due to its size and
complexity.

\subsection{Strong scaling experiments}

%The results of the complexity analysis (equations~\ref{eq:kmer} -~\ref{eq:gap}) are summarized in Tables~\ref{tab:params} and~\ref{tab:summary}. In particular, Table~\ref{tab:params} lists the parameters used in our complexity model and Table~\ref{tab:summary} illustrates the complexity of the main stages in the pipeline.
%
%\begin{tabular}{ | c | c |}
%\hline
% \textbf{Symbol} & \textbf{Parameter Description}  \\ \hline
% $G$ & genome size  \\ \hline
% $d$ & depth of coverage  \\ \hline
% $L$ & read length  \\ \hline
% $k$ & length of $k$-mers \\ \hline
% $\alpha$ & average number of alignments per read\\ \hline
% $\beta$ & width of band in Smith-Waterman\\ \hline
% $\gamma$ & \% of genome not in assembled contigs\\ \hline
%\end{tabular}
%
%\begin{tabular}{ | c | c |}
%\hline
% \textbf{Stage} & \textbf{Complexity}  \\ \hline
% $k$-mer analysis & $3\cdot G\cdot d/L \cdot(L-k+1)$  \\ \hline
% contig generation & $G$  \\ \hline
% sequence alignment & $G \cdot (1+d\cdot \alpha \cdot \beta)$  \\ \hline
% scaffolding & $G\cdot d/L \cdot \alpha + 4G$ \\ \hline
% gap closing & $\gamma G \cdot (1+ d/L \cdot(L-k+1))$ \\ \hline
%\end{tabular}

\begin{figure}[!t]
\centering
\includegraphics[width=0.7\textwidth]{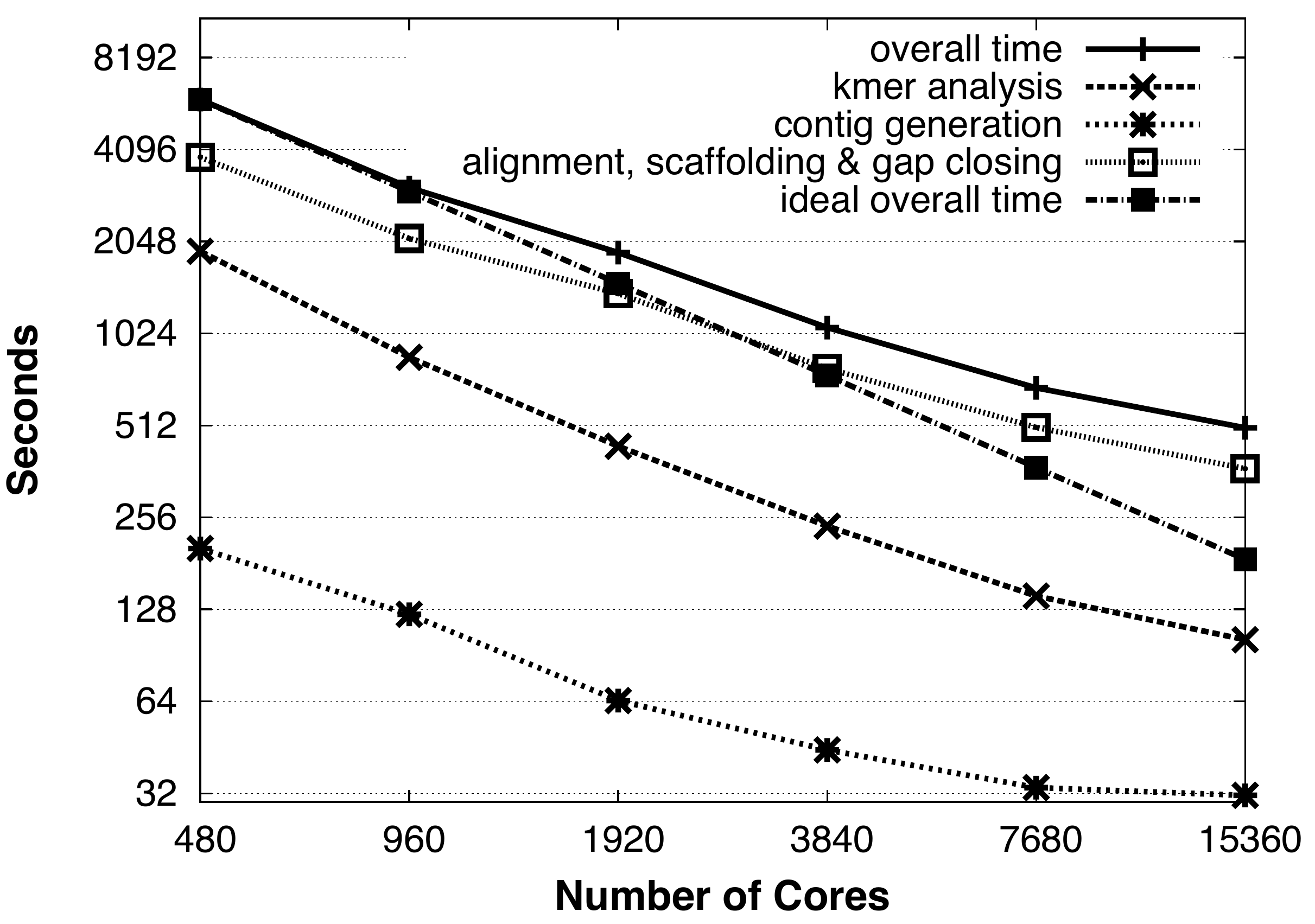}
\caption{End-to-end strong scaling for the human genome. Both axes are in log scale.}
\label{fig:ete1}
\end{figure}
\begin{figure}[!t]
\centering
\includegraphics[width=0.7\textwidth]{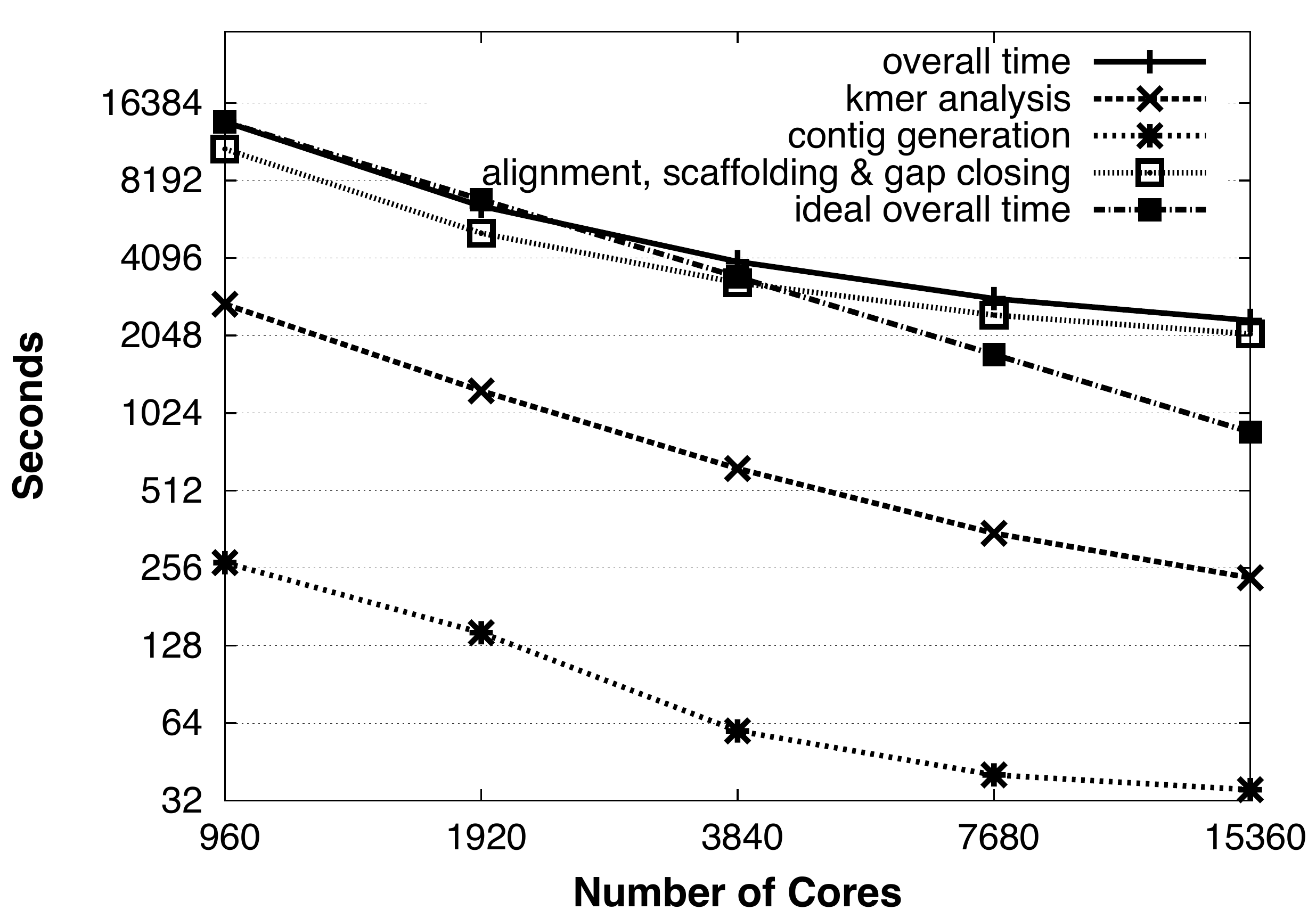}
\caption{End-to-end strong scaling for the wheat genome. Both axes are in log scale.}
\label{fig:ete2}
\end{figure}
Figures~\ref{fig:ete1} and ~\ref{fig:ete2} show the end-to-end strong scaling performance of HipMer (including I/O) with the human and the wheat datasets respectively on the Edison supercomputer. For the human dataset at 15,360 cores we achieve a speedup of
$11.9\times$ over our baseline execution (480 cores). At this extreme
scale the human genome can be assembled from raw reads in just $\approx8.4$ minutes. On the complex wheat dataset, we achieve a speedup up to $5.9\times$
over the baseline of 960 core execution, allowing us to perform the
end-to-end assembly in 39 minutes when using 15,360
cores. In the end-to-end experiments, a significant fraction of the execution time
is spent in parallel sequence alignment, scaffolding and gap closing (e.g.\  $68\%$ for human at 960 cores);
$k$-mer analysis requires less runtime ($28\%$ at 960 cores) and contig generation is the least expensive computational component ($4\%$ at 960 cores).

The $k$-mer analysis and the contig generation steps scale efficiently for both data sets up to 15,360 cores, while the \emph{combined} step of alignment, scaffolding and gap closing exhibits better scaling on the human dataset. Even though the alignment and gap closing modules for the wheat data set exhibit similar scaling to the human test case, the scaffolding step consumes a significantly higher fraction of the overall runtime. There are two main reasons for this behavior. First, the highly repetitive nature of the wheat genome leads to increased fragmentation of the contig generation compared with the human DNA, resulting in contig graphs that are contracted by a smaller fraction. Hence, the serial component of the scaffolding module requires a relatively higher overhead compared with the human dataset. Second, the execution of the wheat pipeline as performed in our previous work~\cite{genomebiology15} requires four rounds of scaffolding with libraries of different insert sizes, resulting in even more overhead within the serial module.

The strong scaling results presented here contradict the conventional wisdom that algorithms with highly irregular accesses (like de novo genome assembly) are prohibitive for distributed memory systems. We showed that as long as the parallel algorithms are highly-scalable and do not exhibit algorithmic/serialization bottlenecks, they perform fewer irregular operations on the critical path as the concurrency increases, therefore decreasing eventually the overall execution time.

\subsection{I/O caching}
\label{sec:caching}

\begin{figure}[t!]
\centering
\includegraphics[width=0.7\textwidth]{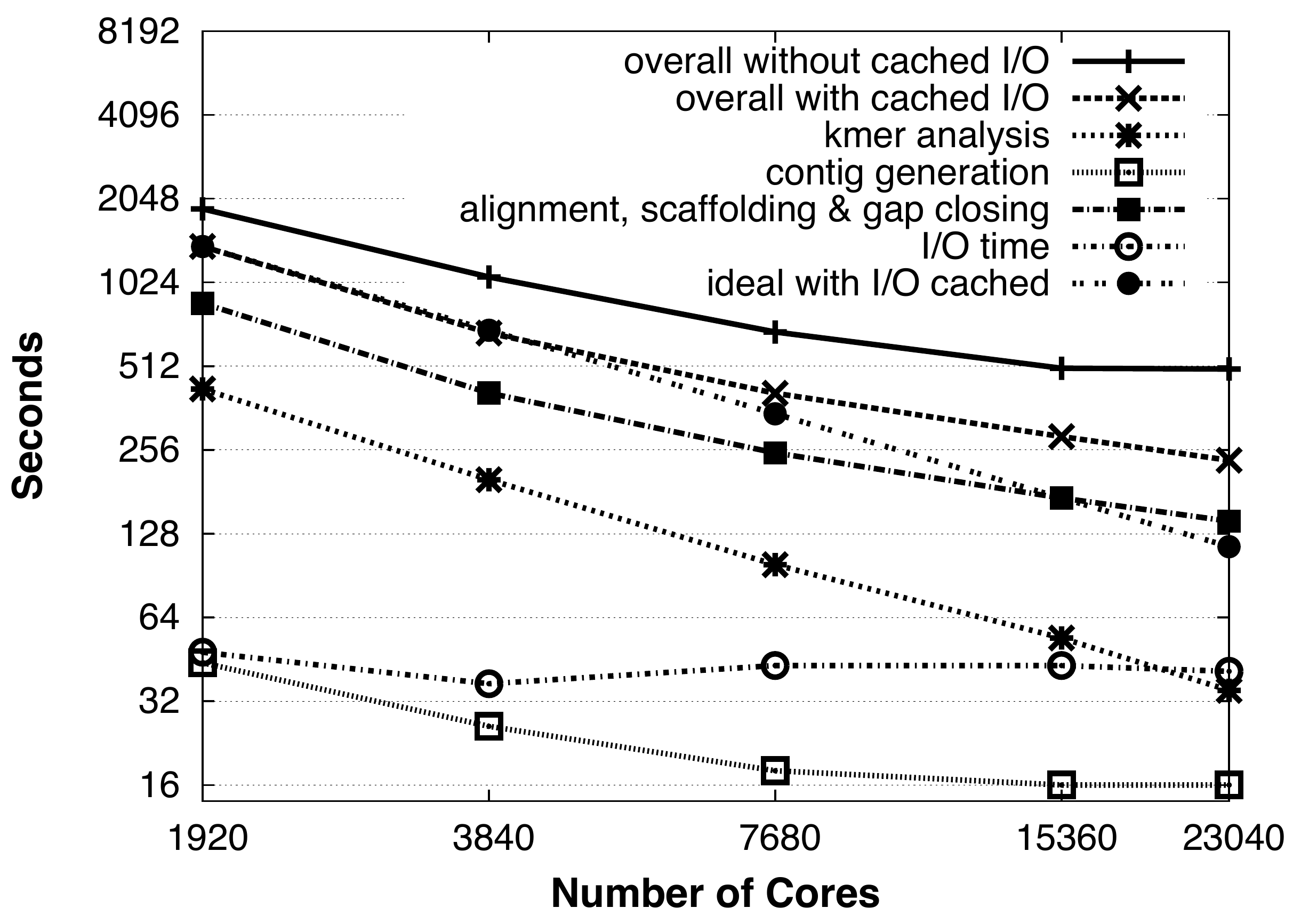}
\caption{Strong scaling of the human data set with I/O caching. Both axes are in logarithmic scale.}
\label{fig:single_strong}
\end{figure}

Our modular design of the pipeline enables flexible configurations that can be adapted appropriately to meet the requirements of each assembly. For instance, one might want to perform multiple rounds of scaffolding to facilitate the assembly of highly repetitive regions or to iterate over the $k$-mer analysis step and contig generation multiple times (with varying $k$ and other parameters) in order to extract information that is latent within a single iteration. These configurations imply that the input read datasets should be loaded multiple times. Even in a typical, single pass execution of the pipeline, the reads constitute the input to multiple stages, namely $k$-mer analysis, alignment and gap closing. Reloading the reads multiple times from the parallel file system, imposes a potential I/O bottleneck for the pipeline. However, at scale we have the unique opportunity to cache the input reads and all the intermediate results onto the aggregate main memory, thus avoiding the excessive I/O and concurrent file system accesses. In order to achieve the I/O caching in a transparent way, we leverage the POSIX shared memory infrastructure and thus all the subsequent input loads are streamed through the main memory.

Figure~\ref{fig:single_strong} shows the end-to-end strong scaling performance of HipMer on the human dataset up to 23,040 Edison cores. We present this experiment in order to highlight the importance of the I/O caching. Note that the baseline concurrency is 1,920 cores; we need at least 80 Edison nodes, each with 24 cores, to fit all the data structures \emph{and} cache the input datasets in memory ($\approx$ 5TB). The line with $\times$ ticks shows the end-to-end execution time \emph{including} the I/O, which is cached in main memory once the input reads are loaded. The ideal strong scaling is illustrated by the line with solid circles. At the concurrency of the 23,040 cores we completely assemble the human genome in 3.91 minutes and obtain a strong scaling efficiency of 48.7\% relative to the baseline of 1,920 cores. In order to illustrate the effectiveness of the I/O caching, we performed the same end-to-end experiments where the input reads are loaded from the Lustre file system five times (solid line). This experiment does not exhibit any scaling from 15,360 to 23,040 cores due to the I/O overhead, thus demonstrating that I/O caching is crucial for scaling to massive concurrencies. At the scale of 23,040 cores, the version with I/O caching is almost $2\times$ faster than the version without this optimization.

The efficiency of the I/O (reading the input reads once) is illustrated by the line with empty circles. We observe that the I/O is almost a flat line across the concurrencies and yields a read bandwidth of $\approx 16$ GB/sec (the theoretical peak of our Lustre file system is 48 GB/sec). With 80 Edison nodes we are able to saturate the available parallelism in the Lustre file system and further increasing the concurrency does not help improve the I/O performance. The lines with {\large{$*$}}, $\square$, $\blacksquare$ ticks show the partial execution time for (i) the $k$-mer analysis, (ii) contig generation and (iii) sequence alignment, scaffolding and gap closing respectively. We conclude that all the components scale up to 23,040 cores and do not impose any scalability impediments. 

\subsection{Performance comparison with other assemblers}
\label{sec:cmp}

To compare the performance of HipMer relative to existing parallel \emph{de
novo} end-to-end genome assemblers we evaluated Ray~\cite{Ray, Ray-Meta}
(version 2.3.0) and ABySS~\cite{abyss} (version 1.3.6) on Edison using
960 cores. Ray required 10 hours and 46 minutes for an end-to-end run on the 
Human dataset.  ABySS, on the other hand, took 13 hours and 26 minutes 
solely to get to the end of contig generation. The subsequent scaffolding
steps are not distributed-memory parallel. At this concurrency on Edison, HipMer is
approximately 13 times faster than Ray and {\em at least} 16 times faster than ABySS.%Both of these assemblers are described in more detail in Chapter~\ref{chap:ten}. 

%Using our HipMer technology enables -- for the first time -- assembly throughput to exceed the capability of all the world's sequencers, thus ushering in a new era of genome analysis. Additionally, HipMer makes it possible to improve assembly quality by running tuning parameter sweeps that were previously prohibitively expense. The combination of high performance sequencing and efficient de novo assembly is the basis for numerous bioinformatic transformations, including advancement of global food security and a new generation of personalized medicine.

\section{Challenges for future architectures}
\label{sec:challenges}

With the advent of exascale computing architectures expected within the next few years, many challenges arise into porting efficiently the HipMer de novo assembly pipeline to larger and more complex systems. In this section we briefly describe these challenges and their implications for highly irregular algorithms, like our de novo assembly pipeline, and the underlying runtime support.

% Mention no sunchronization and more asynchronicity
The architectural trends dictate that the degree of parallelism within the system's node will be increased considerably compared to contemporary supercomputing systems. For instance, the Edison supercomputer (used for our experimental evaluation) is equipped with 24-core nodes, while NERSC's newest supercomputer, named Cori, features Intel Xeon Phi ``Knight's Landing" nodes, each having 68 cores. We expect this trend to hold on the way to exascale. Also, the number of nodes on exascale systems is expected to rise significantly. The combination of the increased number of cores per node and the large number of nodes will yield an unprecedented level of parallelism that should be exploited by the algorithms. In such a massively parallel environment it is crucial to adopt asynchronous algorithmic approaches that do not suffer from load imbalance and system performance fluctuations. The parallel hash table construction and the parallel de Bruijn graph traversal algorithms described in Section~\ref{subsec:contig} are examples of such asynchronous algorithms that do not exhibit synchronization bottlenecks on the critical path of execution. On the other hand, parallel algorithms which rely on bulk synchronous communication will most likely be inadequate for applications with highly irregular accesses. 

% Mention different communicaiton aspects and how they would affect system
In Section~\ref{sec:parallel} we highlighted the different communication patterns that are stressed throughout the HipMer pipeline, namely all-to-all exchanges, irregular lookups and global atomics. Accommodating these communication operations efficiently as the system size increases is critical into porting HipMer to exascale architectures. More specifically, the all-to-all exchange primitives should be mapped efficiently on the underlying network topologies in order to maximize the attainable bandwidth, and ideally should avoid excessive synchronization. Additionally, the operations which are latency-bound like the irregular lookups and the global atomics should exploit efficient protocols and routing algorithms that avoid hot spots on a large-scale system. Furthermore, taking advantage of network capabilities like Remote Direct Memory Access (RDMA) and hardware atomics will play a tremendous role in obtaining low-latency and low-overhead communication primitives. The aforementioned communication optimizations would be preferably applied at the runtime level and therefore could be seamlessly employed at the HipMer application level.

% Mention more hierarchical approaches that do not scale with # of threads
% Mention runtime overheads from "flat" model
All the parallel algorithms in Section~\ref{sec:parallel} are detailed in the context of a flat SPMD execution model, where each UPC thread is mapped onto a compute core of the system. However, the way these UPC threads are instantiated during execution time has implications for the overall performance and the memory footprint of the runtime. For instance, one could use one process per UPC thread; alternatively, one could opt for hierarchical approaches where multiple UPC threads are mapped onto a single process. Both approaches have advantages and disadvantages, but with the arrival of exascale it is imperative to take into account the scale of the systems and re-evaluate the design space. Designs where the runtime's data structures scale in size and complexity quadratically with the number of nodes and/or cores per node are prohibitive. With this in mind, it is more likely that highly optimized hierarchical designs would be suitable for runtimes that target exascale systems. One could additionally adopt analogous hierarchical strategies at the HipMer's application level. However, dealing with this issue upfront at the runtime/communication library level would provide a more robust ecosystem and make the HipMer codebase more portable.

Even though the performance of the HipMer pipeline is mostly dominated by communication and subsequently by the way the communication is orchestrated within the parallel algorithms, it is crucial to optimize the core computations for the underlying architectures. Such computations include mostly string operations (e.g.\ $k$-mer extraction, reverse complementation of sequences, local alignment of sequences, string comparisons) and calculations of hash values. These computations can take advantage of vectorization and hence it is important to leverage vectorized implementations of these core computations throughout the pipeline. This necessity is even more emphasized within the context of the current architectural trends, where the single-thread performance heavily depends on efficient utilization of the vector units.

The process of porting our assembly pipeline to exascale can be tackled on multiple fronts. In the previous paragraphs we explained how some of the key performance factors lie within the UPC runtime level. From this point of view, effective portability of the pipeline is translated into efficient UPC runtime implementation for exascale systems. An additional opportunity to facilitate the porting to exascale systems emerges within the context of the distributed data structures described in Section~\ref{sec:hash}. We could capitalize on the level of abstraction offered by our distributed data structures and their Use Cases (see Table~\ref{tab:opts}) and optimize the core operations of the pipeline at the library level of the data structures. The benefit of such an approach is that the distributed data structure library utilized in HipMer could be specialized for each target system (e.g.\ with appropriate communication optimizations, topology and hierarchical considerations) while the core codebase will remain unmodified. Finally, we highlight that the parallel algorithms in the HipMer pipeline are designed to scale to massive concurrencies and do not exhibit fundamental impediments in porting them to exascale systems.

\section{Related work}
\label{sec:related}

As there are many \emph{de novo} genome assemblers and assessment of the quality of these is well beyond the scope of this chapter, we refer the reader to the work of the Assemblathons I~\cite{assemblathon-brief} and II~\cite{Assemblathon2} as examples of why Meraculous~\cite{meraculous} was chosen to be scaled, optimized and re-implemented as HipMer. For performance comparisons, we primarily refer to parallel assemblers with the potential for strong scaling on large genomes (such as plant, mammalian and metagenomes) using distributed computing or clusters.

Ray~\cite{Ray,Ray-Meta} is an end-to-end parallel \emph{de novo} genome assembler that utilizes MPI and exhibits strong scaling.  It can produce scaffolds directly from raw sequencing reads and produces timing logs for every stage.  One drawback of Ray is the lack of parallel I/O support for reading and writing files. As shown in Section~\ref{sec:performance} Ray is approximately 13$\times$ slower than HipMer for the human data set on 960 cores.

ABySS~\cite{abyss} was the first \emph{de novo} assembler written in MPI that also exhibits strong scaling.  Unfortunately only the first assembly step of contig generation is fully parallelized with MPI and the subsequent scaffolding steps must be performed on a single shared memory node. As shown in Section~\ref{sec:performance} ABySS' contig generation phase is approximately 16$\times$ slower than HipMer's entire end-to-end solution for the human data set on 960 cores.

PASHA~\cite{PASHA} is another partly MPI based de Bruijn graph assembler, though not all steps are fully parallelized as its algorithm, like ABySS, requires a large memory single node for the last scaffolding stages.  The PASHA authors do claim over $2\times$ speedup over ABySS on the same hardware.

YAGA~\cite{jackson2010parallel} is a parallel distributed-memory that is shown to be scalable except for its I/O, but the authors could not obtain a copy of this software to evaluate.  HipMer employs efficient, parallel I/O so is expected to achieve end-to-end performance scalability. Also, the YAGA assembler was designed in an era when the short reads were extremely short and therefore its run-time will be much slower for current high throughout sequencing systems.

SWAP~\cite{SWAP} is a relatively new parallelized MPI based de Bruijn assembler that has been shown to assemble contigs for the human genome and performs strong scaling up to about one thousand cores. However, SWAP does not perform any of the scaffolding steps, and is therefore not an end-to-end \emph{de novo} solution. Additionally, the peak memory usage of SWAP is much higher than HipMer, as it does not leverage Bloom filters. 

% Due to memory constraints, the minimal SWAP assembly of our human dataset on Edison required 1536 cores, and consumed 100 minutes to produce contigs, so HipMer is \emph{at least} 2x faster as an end-to-end scafolding assembler using just 63% (960 cores) of the computational resources that SWAP required.

There are several other shared memory assemblers that produce high quality assemblies, including ALLPATHS-LG~\cite{Allpaths-lg} (pthreads/OpenMP parallel depending on the stage),
%, which imposes certain restrictions on the type of mate-pair inputs it will assemble,
 SOAPdenovo~\cite{li2010novo} (pthreads), DiscovarDenovo~\cite{Discovar} (pthreads) and SPADES~\cite{spades} (pthreads), but unfortunately each of these requires a large memory node and we were unsuccessful at running these experiments using our datasets on a system containing 512GB of RAM due to lack of memory.  This shows the importance of strong scaling distributed memory solutions when assembling large genomes.

\section{Conclusions}
\label{sec:conclusion}

In this chapter we presented HipMer, the first end-to-end highly scalable, high-quality de novo genome assembler, demonstrated to scale efficiently on tens of thousands of cores. HipMer is two orders of magnitude faster than the original Meraculous code and at least an order of magnitude faster than other assemblers, including those with incomplete pipelines and lower quality. Parts of the HipMer pipeline were used in the first whole-genome assembly of the grand-challenge wheat genome~\cite{genomebiology15}. HipMer is so fast, that by using just 17\% of Edison's cores, we could assemble 90 Tbases/day, or all of the 5,400 Tbases in the Sequence Read Archive~\cite{sraweb} in just 2 months. Also, the HipMer technology makes it possible to improve assembly quality by running tuning parameter sweeps that were previously prohibitive in terms of computation. 

%Using the HipMer technology enables -- for the first time -- assembly throughput to exceed the capability of all the world's sequencers, thus ushering in a new era of genome analysis. Additionally, HipMer makes it possible to improve assembly quality by running tuning parameter sweeps that were previously prohibitive in terms of computation. 

%We also presented meta-HipMer, the first scalable, distributed memory metagenome assembler. We showed that meta-HipMer produces assemblies that are competitive or better in quality than those of previous state-of-the-art metagenome assemblers, but at least an order of magnitude faster, because our pipeline can scale efficiently to distributed memory architectures. But most importantly, meta-HipMer is not limited by the concurrency and memory of a single node and thus it can handle multi-terabyte metagenome datasets that other tools are incapable of dealing with. Due to the unforeseen drop in sequencing costs, it is now routine for sequence datasets from a single sample to exceed one terabyte in size and expected to grow by at least an order of magnitude by 2020.

Obtaining this scalable pipeline required several new parallel algorithms and distributed data structures which take advantage of a global address space model of computation on distributed memory hardware, remote atomic memory operations and novel synchronization protocols. Additionally, we developed runtime support to reduce communication cost through dynamic message aggregation, and statistical algorithms that reduced communication through locality aware hashing schemes. We showed that high-performance distributed hash tables, with various optimizations constitute a powerful abstraction for this type of irregular data analysis problems. 

We believe our results will be important both in the application of assembly to health and environmental applications and in providing a conceptual framework for scalable genome analysis algorithms beyond those presented here. The code for HipMer is open source and can be downloaded at: \texttt{https://sourceforge.net/projects/hipmer/}.

\bibliographystyle{plain}
\bibliography{../chapter_genome,../kmercount,../references,../hipmer}

\begin{thebibliography}{10}

\bibitem{sraweb}
Sra database growth.
\newblock \url{http://www.ncbi.nlm.nih.gov/sra/docs/sragrowth/}.
\newblock Accessed: 2016-07-18.

\bibitem{balakrishnan2003looking}
Hari Balakrishnan, M~Frans Kaashoek, David Karger, Robert Morris, and Ion
  Stoica.
\newblock Looking up data in p2p systems.
\newblock {\em Communications of the ACM}, 46(2):43--48, 2003.

\bibitem{spades}
Anton Bankevich, Sergey Nurk, Dmitry Antipov, Alexey~A. Gurevich, Mikhail
  Dvorkin, Alexander~S. Kulikov, Valery~M. Lesin, Sergey~I. Nikolenko, Son
  Pham, Andrey~D. Prjibelski, Alexey~V. Pyshkin, Alexander~V. Sirotkin, Nikolay
  Vyahhi, Glenn Tesler, Max~A. Alekseyev, and Pavel~A. Pevzner.
\newblock {SPAdes}: A new genome assembly algorithm and its applications to
  single-cell sequencing.
\newblock {\em J Comput Biol.}, 19(5):455--477, May 2012.

\bibitem{bloom1970space}
Burton~H Bloom.
\newblock Space/time trade-offs in hash coding with allowable errors.
\newblock {\em Communications of the ACM}, 13(7):422--426, 1970.

\bibitem{Ray}
S{\'e}bastien Boisvert, Fran{\c{c}}ois Laviolette, and Jacques Corbeil.
\newblock Ray: simultaneous assembly of reads from a mix of high-throughput
  sequencing technologies.
\newblock {\em Journal of Computational Biology}, 17(11):1519--1533, 2010.

\bibitem{Ray-Meta}
S{\'e}bastien Boisvert, Fr{\'e}d{\'e}ric Raymond, Él{\'e}nie Godzaridis,
  Fran{\c{c}}ois Laviolette, and Jacques Corbeil.
\newblock Ray meta: scalable de novo metagenome assembly and profiling.
\newblock {\em Genome Biology}, 13(R122), 2012.

\bibitem{gasnet}
Dan Bonachea.
\newblock Gasnet specification, v1.1.
\newblock \url{http://gasnet.lbl.gov/CSD-02-1207.pdf}, 2002.

\bibitem{Assemblathon2}
Keith Bradnam1, Joseph Fass, Anton Alexandrov, et~al.
\newblock Assemblathon 2: evaluating de novo methods of genome assembly in
  three vertebrate species.
\newblock {\em GigaScience}, 2(10), 2013.

\bibitem{shmem}
Barbara Chapman, Tony Curtis, Swaroop Pophale, Stephen Poole, Jeff Kuehn, Chuck
  Koelbel, and Lauren Smith.
\newblock Introducing openshmem: Shmem for the pgas community.
\newblock In {\em Proceedings of the Fourth Conference on Partitioned Global
  Address Space Programming Model}, page~2. ACM, 2010.

\bibitem{Merac2}
JA~Chapman, I~Ho, E~Goltsman, and DS~Rokhsar.
\newblock Meraculous2: fast accurate short-read assembly of large polymorphic
  genomes.
\newblock {\em PLOS}, Submitted, 2016.

\bibitem{genomebiology15}
Jarrod Chapman, Martin Mascher, Aydin Buluc, Kerrie Barry, Evangelos Georganas,
  Adam Session, Veronika Strnadova, Jerry Jenkins, Sunish Sehgal, Leonid
  Oliker, Jeremy Schmutz, Katherine Yelick, Uwe Scholz, Robbie Waugh, Jesse
  Poland, Gary Muehlbauer, Nils Stein, and Daniel Rokhsar.
\newblock A whole-genome shotgun approach for assembling and anchoring the
  hexaploid bread wheat genome.
\newblock {\em Genome Biology}, 16(26), 2015.

\bibitem{meraculous}
Jarrod~A. Chapman, Isaac Ho, Sirisha Sunkara, Shujun Luo, Gary~P. Schroth, and
  Daniel~S. Rokhsar.
\newblock Meraculous: De novo genome assembly with short paired-end reads.
\newblock {\em PLoS ONE}, 6(8):e23501, 08 2011.

\bibitem{expected}
Richard~C Deonier, Simon Tavar{\'e}, and Michael Waterman.
\newblock {\em Computational genome analysis: an introduction}.
\newblock Springer Science \& Business Media, 2005.

\bibitem{mpi3}
James Dinan, Pavan Balaji, Darius Buntinas, David Goodell, William Gropp, and
  Rajeev Thakur.
\newblock An implementation and evaluation of the mpi 3.0 one-sided
  communication interface.
\newblock {\em Concurrency and Computation: Practice and Experience}, 2016.

\bibitem{assemblathon-brief}
Dent Earl, Keith Bradnam, John St~John, Aaron Darling, et~al.
\newblock {Assemblathon 1: a competitive assessment of de novo short read
  assembly methods.}
\newblock {\em Genome research}, 21(12):2224--2241, December 2011.

\bibitem{upc}
Tarek El-Ghazawi and Lauren Smith.
\newblock Upc: unified parallel c.
\newblock In {\em Proceedings of the 2006 ACM/IEEE conference on
  Supercomputing}, page~27. ACM, 2006.

\bibitem{ellis1987concurrency}
Carla~Schlatter Ellis.
\newblock Concurrency in linear hashing.
\newblock {\em ACM Transactions on Database Systems (TODS)}, 12(2):195--217,
  1987.

\bibitem{DMTCSPROCdmAH0110}
P.~Flajolet, E.~Fusy, O.~Gandouet, and F.~Meunier.
\newblock Hyperloglog: the analysis of a near-optimal cardinality estimation
  algorithm.
\newblock {\em DMTCS Proceedings}, 2008.

\bibitem{georganas2016scalable}
Evangelos Georganas.
\newblock {\em Scalable Parallel Algorithms for Genome Analysis}.
\newblock PhD thesis, University of California, Berkeley, 2016.

\bibitem{hipmer}
Evangelos Georganas, Aydin Buluc, Jarrod Chapman, Steven Hofmeyr, Chaitanya
  Aluru, Rob Egan, Leonid Oliker, Daniel Rokhsar, and Katherine Yelick.
\newblock {HipMer}: an extreme-scale de novo genome assembler.
\newblock In {\em Proceedings of the International Conference for High
  Performance Computing, Networking, Storage and Analysis}, page~14. ACM, 2015.

\bibitem{sc14}
Evangelos Georganas, Aydin Buluc, Jarrod Chapman, Leonid Oliker, Daniel
  Rokhsar, and Katherine Yelick.
\newblock Parallel {D}e {B}ruijn {G}raph {C}onstruction and {T}raversal for
  {D}e {N}ovo {G}enome {A}ssembly.
\newblock In {\em Proceedings of the International Conference for High
  Performance Computing, Networking, Storage and Analysis (SC'14)}, 2014.

\bibitem{ipdps15}
Evangelos Georganas, Aydin Buluc, Jarrod Chapman, Leonid Oliker, Daniel
  Rokhsar, and Katherine Yelick.
\newblock mer{A}ligner: {A} {F}ully {P}arallel {S}equence {A}ligner.
\newblock In {\em Proceedings of the IPDPS}, 2015.

\bibitem{gerstenberger2013enabling}
Robert Gerstenberger, Maciej Besta, and Torsten Hoefler.
\newblock Enabling highly-scalable remote memory access programming with mpi-3
  one sided.
\newblock In {\em 2013 SC-International Conference for High Performance
  Computing, Networking, Storage and Analysis (SC)}, pages 1--12. IEEE, 2013.

\bibitem{Allpaths-lg}
S~Gnerre, D~MacCallum, I~andPrzybylski, F~Ribeiro, J~Burton, B~Walker,
  T~Sharpe, G~Hall, T~Shea, S~Sykes, A~Berlin, D~Aird, M~Costello, R~Daza,
  L~Williams, R~Nicol, A~Gnirke, C~Nusbaum, ES~Lander, and DB~Jaffe.
\newblock High-quality draft assemblies of mammalian genomes from massively
  parallel sequence data.
\newblock In {\em Proceedings of the National Academy of Sciences USA}, 2010.

\bibitem{mpi}
William Gropp, Ewing Lusk, Nathan Doss, and Anthony Skjellum.
\newblock A high-performance, portable implementation of the mpi message
  passing interface standard.
\newblock {\em Parallel computing}, 22(6):789--828, 1996.

\bibitem{hopscotch}
Maurice Herlihy, Nir Shavit, and Moran Tzafrir.
\newblock Hopscotch hashing.
\newblock In {\em International Symposium on Distributed Computing}, pages
  350--364. Springer, 2008.

\bibitem{hsu1986concurrent}
Meichun Hsu and Wei-Pang Yang.
\newblock Concurrent operations in extendible hashing.
\newblock In {\em VLDB}, volume~86, pages 25--28, 1986.

\bibitem{jackson2010parallel}
Benjamin~G Jackson, Matthew Regennitter, et~al.
\newblock Parallel de novo assembly of large genomes from high-throughput short
  reads.
\newblock In {\em IPDPS'10}. IEEE, 2010.

\bibitem{Discovar}
David Jaffe.
\newblock Discovar: Assemble genomes and find variants.
\newblock \url{http://www.broadinstitute.org/software/discovar/blog/}, 2014.

\bibitem{Myers1995}
J.D. Kececioglu and E.W. Myers.
\newblock Combinatorial algorithms for dna sequence assembly.
\newblock {\em Algorithmica}, 13:7–51, 1995.

\bibitem{pami}
Sameer Kumar, Amith~R Mamidala, Daniel~A Faraj, Brian Smith, Michael Blocksome,
  Bob Cernohous, Douglas Miller, Jeff Parker, Joseph Ratterman, Philip
  Heidelberger, et~al.
\newblock Pami: A parallel active message interface for the blue gene/q
  supercomputer.
\newblock In {\em Parallel \& Distributed Processing Symposium (IPDPS), 2012
  IEEE 26th International}, pages 763--773. IEEE, 2012.

\bibitem{kumar1990concurrent}
Vijay Kumar.
\newblock Concurrent operations on extendible hashing and its performance.
\newblock {\em Communications of the ACM}, 33(6):681--694, 1990.

\bibitem{li2010novo}
Ruiqiang Li, Hongmei Zhu, et~al.
\newblock De novo assembly of human genomes with massively parallel short read
  sequencing.
\newblock {\em Genome research}, 20(2):265--272, 2010.

\bibitem{liu2013estimation}
Binghang Liu, Yujian Shi, Jianying Yuan, Xuesong Hu, Hao Zhang, Nan Li, Zhenyu
  Li, Yanxiang Chen, Desheng Mu, and Wei Fan.
\newblock Estimation of genomic characteristics by analyzing k-mer frequency in
  de novo genome projects.
\newblock {\em arXiv preprint arXiv:1308.2012}, 2013.

\bibitem{PASHA}
Yongchao Liu, Bertil Schmidt, and Douglas~L Maskell.
\newblock Parallelized short read assembly of large genomes using de {B}ruijn
  graphs.
\newblock {\em BMC bioinformatics}, 12(1):354, 2011.

\bibitem{mayer2014chromosome}
Klaus~FX Mayer, Jane Rogers, Jaroslav Dole{\v{z}}el, Curtis Pozniak, Kellye
  Eversole, Catherine Feuillet, Bikram Gill, Bernd Friebe, Adam~J Lukaszewski,
  Pierre Sourdille, et~al.
\newblock A chromosome-based draft sequence of the hexaploid bread wheat
  (triticum aestivum) genome.
\newblock {\em Science}, 345(6194), 2014.

\bibitem{maynard2012comparing}
Chris Maynard.
\newblock Comparing one-sided communication with mpi, upc and shmem.
\newblock {\em Proceedings of the Cray User Group (CUG)}, 2012, 2012.

\bibitem{melsted2011efficient}
P{\'a}ll Melsted and Jonathan~K Pritchard.
\newblock Efficient counting of k-mers in {DNA} sequences using a bloom filter.
\newblock {\em BMC bioinformatics}, 12(1):333, 2011.

\bibitem{SWAP}
Jintao Meng, Bingqiang Wang, Yanjie Wei, Shengzhong Feng, and Pavan Balaji.
\newblock {SWAP}-assembler: scalable and efficient genome assembly towards
  thousands of cores.
\newblock {\em BMC Bioinformatics}, 15(Suppl 9):S2, 2014.

\bibitem{michael2002high}
Maged~M Michael.
\newblock High performance dynamic lock-free hash tables and list-based sets.
\newblock In {\em Proceedings of the fourteenth annual ACM symposium on
  Parallel algorithms and architectures}, pages 73--82. ACM, 2002.

\bibitem{ratnasamy2001scalable}
Sylvia Ratnasamy, Paul Francis, Mark Handley, Richard Karp, and Scott Shenker.
\newblock {\em A scalable content-addressable network}, volume~31.
\newblock ACM, 2001.

\bibitem{shalev2006split}
Ori Shalev and Nir Shavit.
\newblock Split-ordered lists: Lock-free extensible hash tables.
\newblock {\em Journal of the ACM (JACM)}, 53(3):379--405, 2006.

\bibitem{shun}
Julian Shun and Guy~E Blelloch.
\newblock Phase-concurrent hash tables for determinism.
\newblock In {\em Proceedings of the 26th ACM symposium on Parallelism in
  algorithms and architectures}, pages 96--107. ACM, 2014.

\bibitem{abyss}
Jared~T Simpson, Kim Wong, et~al.
\newblock Abyss: a parallel assembler for short read sequence data.
\newblock {\em Genome research}, 19(6):1117--1123, 2009.

\bibitem{Simpson2015}
JT~Simpson and M~Pop.
\newblock The theory and practice of genome sequence assembly.
\newblock {\em Annu Rev Genomics Hum Genet.}, pages 153--72, 2015.

\bibitem{SW}
Temple~F Smith and Michael~S Waterman.
\newblock Identification of common molecular subsequences.
\newblock {\em Journal of molecular biology}, 147(1):195--197, 1981.

\bibitem{dmapp}
Monika ten Bruggencate and Duncan Roweth.
\newblock Dmapp-an api for one-sided program models on baker systems.
\newblock In {\em Cray User Group Conference}, 2010.

\bibitem{zhang2013survey}
Hao Zhang, Yonggang Wen, Haiyong Xie, and Nenghai Yu.
\newblock A survey on distributed hash table (dht): Theory, platforms, and
  applications, 2013.

\bibitem{loblolly_Zimin}
Aleksey Zimin, Kristian~A. Stevens, Marc~W. Crepeau, Anne Holtz-Morris, Maxim
  Koriabine, Guillaume Mar\c~cais, Daniela Puiu, Michael Roberts, Jill~L.
  Wergrzyn, Pieter~J. de~Jong, David~B. Neale, Steven~L. Salzbert, James~A.
  Yorke, and Charles~H. Langley.
\newblock Sequencing and assembly of the 22-gb loblolly pine genome.
\newblock {\em Genetics}, 196(3):875--90, Mar 2014.

\end{thebibliography}

\printindex

\end{document}